\documentclass[aps, pra, preprint, amsmath]{revtex4-1}

\usepackage[T1]{fontenc}
\usepackage[utf8]{inputenc}
\usepackage{amsmath, amssymb, amsfonts, bm, amsthm, braket}
\usepackage{rotating, graphicx}
\usepackage{color}
\usepackage{xr}
\usepackage{hyperref} % Must always be last

\newcommand{\half}{\frac{1}{2}}
\newcommand{\kket}[1]{| #1 \rangle \rangle}
\newcommand{\bbra}[1]{\langle\langle #1|}
\newcommand{\bbrakket}[2]{\langle \langle #1|#2\rangle\rangle}
\newcommand{\Liou}{\hat{\hat{\mathcal{L}}}}

\newcommand{\Pp}{\mathcal{P}}
\newcommand{\Qq}{\mathcal{Q}}

\DeclareMathOperator{\Tr}{Tr}

\begin{document}
\date{\today}
\flushbottom

\title{Generalized Adiabatic Theorems: Quantum Systems Driven by Modulated Time-Varying Fields}
\author{Amro Dodin}
\affiliation{Department of Chemistry, Massachusetts Institute of Technology, Cambridge, Massachusetts, 02139, USA}
\email{adodin@mit.edu}

\author{Paul Brumer}
\affiliation{Chemical Physics Theory Group, Department of Chemistry and Center for Quantum Information and Quantum Control, University of Toronto, 
Toronto, Ontario, M5S 3H6, Canada}
\email{Paul.Brumer@utoronto.ca}

\begin{abstract}
  We present generalized adiabatic theorems for closed and open quantum systems that can be applied to slow modulations of rapidly varying 
  fields, such as oscillatory fields that occur in optical experiments and light induced processes.
  The generalized adiabatic theorems show that a sufficiently slow modulation conserves the
  dynamical  modes of time dependent reference Hamiltonians.  
  In the limiting case of modulations of static fields,  the standard adiabatic theorems are recovered.
  Applying these results to periodic fields shows that they remain in Floquet states rather than in energy eigenstates.
  More generally, these adiabatic theorems can be applied to transformations of arbitrary time-dependent fields, by accounting for the rapidly 
  varying part of the field through the dynamical normal modes, and treating the slow modulation adiabatically.
  As examples, we apply the generalized theorem to (a) predict the dynamics of a two level system driven by a frequency modulated resonant 
  oscillation, a pathological situation beyond the applicability of earlier results, and (b) to show that open quantum systems driven by slowly 
  turned-on incoherent light, such as biomolecules under natural illumination conditions, can only display coherences that survive in the steady 
  state.
\end{abstract}

\maketitle

\section{Introduction}
\label{sec:intro}
% Motivate the Problem

Quantum systems driven by slowly varying external fields play an important role in atomic, molecular and optical physics.
The ubiquity of these processes has led to sustained interest in  powerful adiabatic theorems (AT's) that characterize their dynamics 
\cite{born_beweis_1928,kato_adiabatic_1950,sarandy_adiabatic_2005,messiah_quantum_2017}.
The best known AT states that a system initialized in an energy eigenstate will remain in that state when driven by a slowly varying field.
Adiabatic processes have found far reaching utility in quantum dynamics \cite{landau_zur_1932,zener_non-adiabatic_1932} and in the development of 
Adiabatic Quantum Computing (AQC) where they are used to reliably realize quantum state transformations 
\cite{albash_adiabatic_2018,barends_digitized_2016}.
Moreover, adiabaticity conditions bound the allowable speed of state transformations, setting the time cost of AQC algorithms and spurring efforts 
to find shortcuts to adiabaticity \cite{torrontegui_energy_2017,del_campo_shortcuts_2013,torrontegui_chapter_2013,campo_shortcuts_2012}.
However, ATs fail for oscillating fields \cite{amin_consistency_2009,du_experimental_2008,tong_quantitative_2005,marzlin_inconsistency_2004}, 
since these fields can induce transitions between energy eigenstates, precisely the opposite of what is required in traditional adiabatic theorems.
Hence traditional ATs can not, in general, be applied to light-induced processes in either isolated or open quantum systems.

Adiabatic dynamics have been successfully deployed in experimental settings through Adiabatic Rapid Passage (ARP)\cite{malinovsky_general_2001} 
and Stimulated Raman Adiabatic Passage (STIRAP) \cite{vitanov_stimulated_2017,bergmann_perspective_2015} techniques.
  In these experiments, the adiabatic theorem provides a robust mechanism for high fidelity state preparation that is not sensitive to slight 
  perturbations in the system properties and has been used, for example, to study molecular reaction dynamics \cite{kaufmann_dissociative_2001} 
  and in the preparation of ultra-cold molecules \cite{kulz_dissociative_1996,aikawa_coherent_2010}.
  However, the domain of these techniques has  been limited by the applicability of the AT, limiting the forms of driving fields and 
  transformation times that can be realized.
  By relaxing key restrictions imposed by the traditional AT, the AMT introduced below significantly broadens the range of experimental 
  techniques, unlocking the potential for faster, more flexible state preparation and system dynamics.

   Numerous scenarios do not admit traditional adiabatic approaches.  Natural light-induced excitation of biological molecules (e.g., in 
   photosynthesis or vision) provides a particularly important example of a challenge to standard ATs insofar as it involves incoherent excitation 
   of open quantum systems with turn-on times that are exceedingly long compared to the time scale of molecular dynamics.
For example, human blinking, the turn-on time of light in vision,  occurs over 0.2 sec as compared to the timescale of  molecular isomerization 
induced by incident light in vision which can, under pulsed laser excitation,  occur in 50 fs or less.
Thus, challenging questions related to the creation of molecular coherences in such natural processes requires an AT adapted to oscillating incoherent 
fields in open quantum systems \cite{dodin_light-induced_2019,dodin_coherent_2016,cheng_dynamics_2009,chenu_coherence_2015}.

In this paper we derive generalized AT's, termed Adiabatic Modulation Theorems (AMTs), for open and closed quantum systems that apply to 
oscillatory and other rapidly varying fields, extending adiabaticity conditions to optically driven processes and providing faster pathways to 
adiabaticity, with the potential to accelerate AQC.
Moreover, we show that the adiabatic transformation time is limited not by the energy gaps in the systems but rather by the frequency differences 
of the instantaneous normal modes, that are often easier to manipulate.
These results exploit the fact that many complicated fields of interest are time-dependent \textit{modulations} (e.g., frequency and amplitude 
modulated oscillations) of simpler fields that have well understood dynamics.

The AMT's derived below show that a system subjected to a  sufficiently slow modulation of a time-dependent field conserves 
the dynamical  modes of time-dependent reference Hamiltonians.
These theorems contain traditional ATs in the limit of modulations of static fields, and generalize easily to modulations of periodic fields, 
which are shown to preserve Floquet states rather than energy eigenstates.
Moreover, we show that the adiabatic transformation time is limited not by the energy gaps in the system but rather by the frequency differences 
of the instantaneous normal modes, that are often easier to manipulate.
{Remarkably, these constructions allow for the design of experimental techniques that go beyond the preparation of time-independent states and 
allow for the preparation of specific \textit{dynamics} using the well developed intuition of adiabatic theory.}

The AMT theorems are widely applicable and two significant examples are provided below.
In the isolated system example, we show how the theorem allows for control of dynamics on the entire Bloch sphere in a two level qubit system.
In applications to open systems, the theorem resolves a longstanding controversy regarding the role of light-induced coherent oscillations in 
biophysical processes.
The generality of the theorems ensure applications to a wide variety of alternative systems.

This paper is organized as follows: Sect. \ref{sec:AMT} proves the adiabatic modulation theorems for both isolated and open systems.
Sample applications, designed to demonstrate the theorems and their applications are provided in Sect. \ref{sec:examples}.

\section{The Adiabatic Modulation Theorems}
\label{sec:AMT}
\subsection{Isolated Systems}
\label{sec:isolated}

Consider a family of time-dependent reference Hamiltonians $\hat{H}_0(t;\lambda)$ indexed by the parameter $\lambda$ (e.g. $\lambda$ may be an 
amplitude $\hat{H}_0(t;\lambda)= \lambda \hat{h}_0(t)$ or frequency $\hat{H}_0(t;\lambda)=\exp(-i\lambda t)\hat{h}_0+h.c.$).
For each value of $\lambda$, setting $\hbar=1$, we define the instantaneous normal modes as
\begin{equation}
  \label{eq:floquettrial}
  \ket{\psi_n(t; \lambda)} \equiv e^{-i\theta_n(t;\lambda)}\ket{n(t; \lambda)},
\end{equation}
where $\theta_n(t;\lambda) \equiv \int^t_0 ds \epsilon_n(s; \lambda)$ is the dynamical phase, $\epsilon_n(t; \lambda)$ is an instantaneous 
quasienergy and the $\ket{n(t;\lambda)}$ is the instantaneous normal mode of $\hat{H}_0(t;\lambda)$.

These normal modes and corresponding quasienergies are found by solving the eigenproblem
\begin{equation}
  \label{eq:FloquetCondition}
  \hat{H}_F(t; \lambda) \ket{n(t; \lambda)} = \epsilon_n(t; \lambda)\ket{n(t; \lambda)},
\end{equation}
where $\hat{H}_F(t; \lambda) \equiv \hat{H}_0(t;\lambda) - i\partial/\partial t$.
That is, the  normal modes are particular solutions to the Time-Dependent Schr\"{o}dinger Equation (TDSE),
\begin{equation}
  \label{eq:TDSE}
  i\frac{\partial}{\partial t} \ket{\Psi(t)} = \hat{H}(t)\ket{\Psi(t)},
\end{equation}
and are well understood for many driving fields.
For example, for time-independent Hamiltonians, the normal modes of static Hamiltonians are energy eigenstates, while for periodic Hamiltonians 
they are Floquet states.

A modulation of a time-dependent Hamiltonian, $\hat{H}_0(t;\lambda)$, is a transformation that varies $\lambda\to\lambda_t$ over a time interval 
$[0, \tau]$ through a modulation protocol $\Lambda \equiv \lbrace \lambda_t|t\in[0,\tau]\rbrace$.
The resultant modulated field  $ \hat{\widetilde{H}}_\Lambda(t)\equiv \hat{H}_0(t;\lambda_t)$ sweeps through the Hamiltonian family defined above 
(e.g. a field with a time-dependent amplitude for $\hat{H}_0 = \lambda h_0(t)$).
Note, as an aside, that this is an operator generalization of modulations in signal processing and encoding theory \cite{byrne_signal_1993}.
Generally, the modulated Hamiltonian can be significantly more complicated than $H_0$ since the modulation function may be highly nonlinear and 
the modulation parameter can vary non-trivially with time.

The dynamics induced by $\hat{\widetilde{H}}_\Lambda(t)$ can now be expressed in terms of instantaneous normal modes of $\hat{H}_0$, Eq. 
(\ref{eq:floquettrial}), to give
\begin{equation}
  \label{eq:trialform}
  \ket{\Psi(t)} \equiv \sum_n c_n(t)e^{-i\theta_n(t;\Lambda)}\ket{n(t; \lambda_t)}
\end{equation}
where $\theta_n (t;\Lambda) \equiv \int^t_0 ds \epsilon_n(s; \lambda_s)/\hbar$.
The instantaneous state of the system can always be expressed in this form since $\hat{H}_F(t;\lambda_t)$ is always Hermitian and therefore has a 
basis of eigenstates $\lbrace \ket{n(t; \Lambda_t)} \rbrace$ that can be used to expand $\ket{\Psi(t)}$ with coefficient $a_n(t) \equiv 
\braket{n(t;\lambda_t)|\Psi(t)}= c_n(t) \exp(-i\theta_n(t))$ \footnote{The modified Hamiltonian is defined using a partial time derivative and 
therefore neglects the implicit time-dependence of $\lambda_t$.}.

This approach mirrors that used to derive traditional ATs, but replacing the eigenstates with normal modes and energies with quasienergies.
We therefore proceed through a similar path to derive adiabaticity conditions under which a system initialized in a normal mode of 
$\hat{H}_0(0;\lambda_0)$ will remain in that normal mode at all times.
Mathematically, this occurs when $c_n(t)$ are decoupled in the TDSE, Eq. (\ref{eq:TDSE}).
Substituting Eq. (\ref{eq:trialform}) into Eq. (\ref{eq:TDSE}) we find
\begin{equation}
  \label{eq:coefficients}
  \sum_n \left(\dot{c}_n(t)\ket{n(t; \lambda_t)} + c_n(t) \dot{\lambda}_t\frac{\partial}{\partial \lambda_t} \ket{n(t; 
  \lambda_t)}\right)e^{-i\theta_n(t;\Lambda)}= 0,
\end{equation}
where we have used the identity  $d\theta_n(t;\Lambda)/dt = \epsilon_n(t;\lambda_t)$ and Eq. (\ref{eq:FloquetCondition}) to cancel terms 
proportional to $\epsilon_n(t; \lambda_t)$.
Projecting onto an $\ket{m(t;\lambda_t)}$, yields equations of motion for the coefficients,
\begin{equation}
  \label{eq:EoM}
  \dot{c}_m(t) = -c_m(t)\dot{\lambda}_t\bra{m} \frac{\partial}{\partial\lambda_t}\ket{n} +\sum_{n \neq m} 
  \frac{\dot{\lambda}_te^{-i\theta_{nm}}\braket{m|\frac{\partial \hat{\widetilde{H}}}{\partial\lambda}|n}}{\epsilon_m - \epsilon_n},
\end{equation}
where we have suppressed the $t$ and $\lambda_t$ dependence for brevity, and $\theta_{nm} \equiv \theta_n- \theta_m$.
To obtain Eq. (\ref{eq:EoM}), Eq. (\ref{eq:FloquetCondition}) was differentiated with respect to $\lambda_t$ to obtain $\braket{m 
|\partial/\partial\lambda_t|n} = \braket{m|\partial\hat{\widetilde{H}}/\partial \lambda_t|n}/(\epsilon_n - \epsilon_m)$.

Equation (\ref{eq:EoM}) takes the same form as the original adiabatic theorem with cross-coupling between normal modes scaled by a rapidly 
oscillating term \cite{sarandy_adiabatic_2005}.
As a result normal modes evolve independently of one another when the following condition is satisfied:
\begin{equation}
  \label{eq:closedadiab}
  \max_{0\leq t \leq \tau}\left|\frac{\dot{\lambda}_t\braket{m|\frac{\partial 
  \hat{H}(t;\lambda_t)}{\partial\lambda_t}|n}}{\epsilon_{nm}(t;\lambda_t)}\right| \ll \min_{0 \leq t \leq 
  \tau}\left|\epsilon_{nm}(t;\lambda_t)\right|
\end{equation}
where $\epsilon_{nm}(t;\lambda_t) \equiv \epsilon_n(t;\lambda_t) - \epsilon_m(t; \lambda_t)$.
\textit{Consequently, if $\lambda_t$ changes much more slowly than the difference in dynamical mode frequencies, the slow modulation will not cross 
couple the normal modes of the modulated Hamiltonian.}

Equation (\ref{eq:closedadiab}) has several important features.
First, if $\hat{H}_0(t;\lambda)$ are $t$-independent, then their normal modes are energy eigenstates and all Hamiltonian time-dependence is 
contained in the modulation.
In this case, the AMT reduces to the traditional AT, simply relabeling the time variable as $\lambda$.
The AMT then extends previous AT's by recognizing them as statements about how time-dependent transformations of Hamiltonian fields impact 
dynamical modes of the TDSE.
These coincide with eigenstates for static Hamiltonians but the same insight can be applied to any field with well-understood dynamics.
For example, if $\hat{H}_0(t;\lambda)$ are periodic, Eq. (\ref{eq:closedadiab}) states that slow modulations do not couple Floquet states.
One key distinction of the AMT is that its speed limits are set by the differences in normal mode frequencies which are often easier to 
manipulate than energy gaps between eigenstates, allowing for faster adiabatic transformations.  A numerical example is discussed in Sect. \ref{Rabi}.

\subsection{Open Systems}
\label{sec:open}

Consider now an adiabatic theorem for open quantum systems.
To prove this theorem we take an approach inspired by Ref. \cite{sarandy_adiabatic_2005}.
The dynamics of an open quantum system are governed by the Liouville von-Neumann (LvN) equation
\begin{equation}
  \label{eq:LvN}
  \frac{d}{dt}\kket{\rho} = \Liou(t) \kket{\rho}
\end{equation}
in the time-convolutionless form \cite{breuer_theory_2007,alicki_quantum_2007,blum_density_2012}.
The double-ket notation $\kket{\cdot}$ is indicates an operator Liouville space with a trace inner product $\bbrakket{A}{B} \equiv 
\Tr\lbrace\hat{A}^\dagger \hat{B}\rbrace$.
By analogy with the isolated system we define modulations by starting with a family of time-dependent Liouvillians $\Liou_0(t;\lambda)$ and 
allowing the modulation parameter to vary over time to give $\Liou_\Lambda(t) \equiv \Liou_0(t;\lambda_t)$.
A modified Liouvillian can then be defined by $\Liou_F(t;\lambda) \equiv \Liou(t;\lambda) - \partial/\partial t$.

The LvN equation appears similar to the TDSE [Eq. (\ref{eq:TDSE})], suggesting that we may be able to apply the same analysis, but replacing 
Hamiltonians with Liouvillians.
However, the Liouvillian and it's corresponding modified operator are completely positive but not necessarily Hermitian \cite{alicki_quantum_2007} 
and therefore require more care in defining their normal modes.
A given $M\times M$ modified Liouvillian, $\Liou_F$, has an incomplete set of $N\leq M$ left and right eigenvectors $\bbra{\Qq_\alpha(t)}$ and 
$\kket{\Pp_\alpha (t)}$.
Associated with each eigenvector is a Jordan Block comprised of $n_\alpha$ generalized eigenvectors that combine to give an orthonormal basis of 
Liouville space defined by the generalized eigenproblem
\begin{subequations}
  \label{eqs:jordan}
  \begin{equation}
    \label{eq:jordanleft}
    \bbra{\Qq_\alpha^{(j)}}\Liou_F = \bbra{\Qq_\alpha^{(j)}}\chi_\alpha +\bbra{\Qq_\alpha^{(j+1)}}
  \end{equation}
  \begin{equation}
    \label{eq:jordanright}
    \Liou_F\kket{\Pp_\alpha^{(j)}} = \chi_\alpha \kket{\Pp_\alpha^{(j)}} + \kket{\Pp_\alpha^{(j-1)}}
  \end{equation}
\end{subequations}
where $\bbra{\Qq_\alpha^{(n_\alpha-1)}}\equiv \bbra{\Qq_\alpha}$,$\bbra{\Qq_\alpha^{(n_\alpha)}}\equiv \bbra{0}$, $\kket{\Pp_\alpha^{(0)}} \equiv 
\kket{\Pp_\alpha}$, and $\kket{\Pp_\alpha^{(-1)}}\equiv\kket{0}$,
where $\kket{0}$ is the zero operator.
Dynamically, the LvN Equation does not cross couple the Jordan blocks to one another, but can lead to cross coupling within states of the same 
Jordan block, thereby defining decoupled dynamical subspace.
(For a concise summary of Jordan canonical form in the context of quantum adiabatic theorems, see ref. \cite{sarandy_adiabatic_2005}.)

Motivated by the isolated systems derivation, we consider the modified Liouvillian and aim to expand the dynamics of the open system in terms of 
its instantaneous generalized eigenstates.
These are given by the (right) generalized equation
\begin{equation}
  \label{eq:floquetequationopen}
  \Liou_F(t; \lambda_t) \kket{\Pp_\alpha^{(j)}} = \chi_\alpha  \kket{\Pp_\alpha^{(j)}} +  \kket{\Pp_\alpha^{(j-1)}}.
\end{equation}
The left generalized eigenstates can be similarly defined as the right eigenstates of $\Liou_F^\dagger$.

Allowing the modulation parameter to change with time, the instantaneous state of the system can be expanded as a superposition over the  
generalized eigenstates in the form
\begin{equation}
  \label{eq:openfloquettrial}
  \kket{\rho(t)} \equiv \sum_{\beta=1}^N \sum_{j=0}^{n_\beta-1}r_\beta^{(j)}(t) \kket{\Pp_\beta^{(j)}(t; \lambda_t)}
\end{equation}
where $r_\beta^{(j)}(t) \equiv \bbrakket{\Qq_\beta^{(j)}(t; \lambda_t)}{\rho(t)}$ is a complex valued expansion coefficient.
Similarly, to Eq. (\ref{eq:trialform}), any density operator $\kket{\rho}$ can be expressed in this form since the generalized eigenbasis that 
generates Jordan canonical form is complete and orthonormal.

Substituting the trial form of Eq. (\ref{eq:openfloquettrial}) into the time convolutionless Liouville-von Neumann equation [Eq. (\ref{eq:LvN})] 
and projecting onto the left generalized eigenstate $\bbra{\Qq_\alpha^{(i)}}$, gives the following equation of motion for the expansion 
coefficients:
\begin{equation}
  \label{eq:coefficientsOpen}
  \dot{r}_\alpha^{(i)}= \chi_\alpha r_\alpha^{(i)} +r_\alpha^{(i+1)} - \sum_{\beta=1}^N\sum_{j=0}^{n_\beta-1}\dot{\lambda}_tr_\beta^{(j)} 
  \bbra{\Qq_\alpha^{(i)}} \frac{\partial}{\partial \lambda_t} \kket{\Pp_\beta^{(j)}}
\end{equation}
where we have used the Floquet Equation (\ref{eq:floquetequationopen}) and the orthonormality condition of generalized eigenstates.
By convention, we take $r_\alpha^{(n_\alpha)}=0$.

Differentiating  Eq. (\ref{eq:floquetequationopen}) for some right eigenstate $\kket{\Pp_\beta^{(j)}}$
with respect to $\lambda_t$ at fixed $t$ and projecting onto left eigenstate $\bbra{\Qq_\alpha^{(i)}}$ with $\alpha \neq\beta$ gives
\begin{equation}
  \label{eq:proj-deriv}
  \bbra{\Qq_\alpha^{(i)}}\frac{\partial\Liou}{\partial \lambda_t}\kket{\Pp_\beta^{(j)}} + \bbra{\Qq_\alpha^{(i)}}\Liou \frac{\partial}{\partial \lambda_t} \kket{\Pp_\beta^{(j)}} = \frac{\partial \chi_\beta}{\partial \lambda_t}\bbrakket{\Qq_\alpha^{(i)}}{\Pp_\beta^{(j)}} + \chi_\beta\bbrakket{\Qq_\alpha^{(i)}}{\dot{\Pp}_\beta^{(j)}}  + \bbrakket{\Qq_\alpha^{(i)}}{\dot{\Pp}_\beta^{(j-1)}}.
\end{equation}
This expression can be simplified by first noting that for $\alpha \neq \beta$ orthonormality of the basis eliminates the $\bbrakket{\Qq_\alpha^{(i)}}{\Pp_\beta^{(j)}}$ term on the right hand side of Eq. (\ref{eq:proj-deriv}).
Equation (\ref{eq:jordanleft}) can then be used to expand the second term, $\bbra{\Qq_\alpha^{(i)}}\Liou ({\partial}/{\partial \lambda_t}) \kket{\Pp_\beta^{(j)}}$ yielding a recursive expression for the projected change of the normal modes.
\begin{equation}
  \label{eq:recursive}
  \begin{split}
    \bbra{\Qq_\alpha^{(i)}} \frac{\partial}{\partial\lambda_t}\kket{\Pp_\beta^{(j)}} &= \left(\frac{1}{\chi_{\beta\alpha}}
    (\bbra{\Qq_\alpha^{(i)}}\frac{\partial\Liou}{\partial\lambda_t}\kket{\Pp_\beta^{(j)}} + 
    \bbra{\Qq_\alpha^{(i+1)}}\frac{\partial}{\partial\lambda_t}\kket{\Pp_\beta^{(j)}}\right.\\
    & \left. -  \bbra{\Qq_\alpha^{(i)}}\frac{\partial}{\partial\lambda_t}\kket{\Pp_\beta^{(j-1)}}\right),
  \end{split}
\end{equation}
where $\chi_{\beta\alpha} \equiv \chi_\beta - \chi_\alpha$.
Iterating recursively through Eq. (\ref{eq:recursive}), the change in the normal modes can be related to the change in the modulated Liouvillian 
by
\begin{equation}
  \label{eq:overlapchange}
  \bbra{\Qq_\alpha^{i}} \frac{\partial}{\partial\lambda}\kket{\Pp_\beta^{(j)}} = \sum_{p=1}^{n_\alpha 
  -i}\sum_{k_1=0}^{j-S_0}\cdot\cdot\cdot\sum_{k_p=0}^{j-S_p}\frac{\bbra{\Qq_\alpha^{(i+p-1)}}\frac{\partial\Liou}{\partial\lambda}\kket{\Pp_\beta^{(j-S_p)}}}{(-1)^{S_p}\chi_{\beta\alpha}^{p+S_p}}
\end{equation}
where $S_q \equiv \sum_{s=1}^q k_s$, and $k_s$ are the summation indexes over states in the Jordan Block.

Finally, by substituting Eq. (\ref{eq:overlapchange}) into Eq. (\ref{eq:coefficientsOpen})
and considering only the terms that couple non-degenerate Jordan blocks $\alpha\neq\beta$, we obtain \textit{an adiabaticity condition for slowly 
modulated open quantum systems:}
\begin{equation}
  \label{eq:openadibatic}
  \begin{split}
    \max_{0\leq t \leq T}&\left|\sum_{p=1}^{n_\alpha 
    -i}\sum_{k_1=0}^{j-S_0}\cdot\cdot\cdot\sum_{k_p=0}^{j-S_p}\frac{\bbra{\Qq_\alpha^{(i+p-1)}}\frac{\partial\Liou}{\partial\lambda}\kket{\Pp_\beta^{(j-S_p)}}}{(-1)^{S_p}\chi_{\beta\alpha}^{p+S_p}}\right| 
    \\
    & \ll \min_{0\leq t \leq T} |\dot{\lambda}_t^{-1}|.
    \end{split}
\end{equation}
This expression is admittedly unwieldy but provides a completely general criteria for open system
adiabaticity.
A number of simpler but less tight bounds can be obtained for the adiabaticity condition
by extending the approach discussed in \cite{sarandy_adiabatic_2005} that treated constant Liouvillian.
The simplest of these results states that if $\dot{\lambda}_t \to 0$,  e.g., in Eq. (\ref{eq:coefficientsOpen}),
then the open quantum system undergoes adiabatic dynamics.
Moreover, similarly to the closed system adiabatic theorem, Eq. (\ref{eq:openadibatic}) can be used to derive the
earlier non-modulated adiabatic theorem of Ref.  \cite{sarandy_adiabatic_2005}
by assuming a constant  $\Liou_0(t)$.

 A significant example of the open system AMT is provided in Sect. \ref{sec:ex-open} where the  theorem is applied to the
slowly turned-on incoherent (e.g., solar radiation) excitation of molecular systems.

\section{Computational Examples}
\label{sec:examples}

\subsection{Isolated Systems}
\subsubsection{Rabi type system}
\label{Rabi}

In this section, we consider a two-level system (TLS) driven by a frequency modulated oscillatory field.
We begin by defining an extension of the Rabi model, the family of Hamiltonians,
\begin{equation}
   \label{eq:RabiHam}
   \hat{H}_0(t;\omega) =
   \begin{pmatrix}
   -{\Delta_0} & V e^{i\omega t} \\
   V^* e^{-i\omega t} & {\Delta_0}
   \end{pmatrix},
\end{equation}
that characterize driving by a field with frequency $\omega$.
Here $\Delta_0$ is the energy difference between the states $\ket{0}$ and $\ket{1}$, the coupling coefficient is given by $V = 
\bm{E}\cdot\bm{\mu_{01}}$, $\bm{E}$ is the electric field vector driving the system, $\bm{\mu_{01}}$ is the transition dipole moment between the 
two states and we have set $\hbar=1$.
This family of Hamiltonians comprises the standard Rabi model with well understood dynamics for all values of $\omega$.

It is useful to briefly review the dynamics induced by Eq. (\ref{eq:RabiHam}) from the perspective of Floquet's theorem to highlight the normal modes 
that play an important role in the generalized adiabatic theorem.  Since $\hat{H}_0(t)$ is periodic with period $T=2\pi/\omega$, it's dynamical normal modes are two $T$-periodic Floquet states, 
$\ket{\pm(t;\omega)}$ with time-independent quasienergies $\epsilon_\pm$.
These states are eigenstates of the Floquet Hamiltonian, satisfying
\begin{subequations}
  \label{eqs:floquet}
    \begin{equation}
    \label{eq:floquetCondition}
    \hat{H}_F(t;\omega)\ket{\pm(t;\omega)} = \epsilon_\pm \ket{\pm(t;\omega)}
  \end{equation}
  \begin{equation}
    \label{eq:floquetHam}
    \hat{H}_F(t;\omega)\equiv \hat{H}_0(t;\omega) - i\frac{\partial}{\partial t}
  \end{equation}.
\end{subequations}
For the Hamiltonian in Eq. (\ref{eq:RabiHam}), these Floquet states and quasi energies are given by
\begin{subequations}
  \label{eqs:RabiSol}
  \begin{equation}
    \label{eq:Rabi+State}
    \ket{+(t;\omega)} =
    \begin{pmatrix}
      \sin\frac{\theta}{2} e^{i\frac{\omega}{2}t} \\
      \cos\frac{\theta}{2} e^{-i\phi}e^{-i\frac{\omega}{2}t}
    \end{pmatrix}
  \end{equation}
  \begin{equation}
    \label{eq:Rabi-State}
    \ket{-(t;\omega)} =
    \begin{pmatrix}
      \cos\frac{\theta}{2}  e^{i\frac{\omega}{2}t} \\
      \sin\frac{\theta}{2} e^{-i\phi}e^{-i\frac{\omega}{2}t}
    \end{pmatrix}
  \end{equation}
  \begin{equation}
    \label{eq:RabiEnergies}
    \epsilon_\pm = \pm\Omega \equiv \pm \sqrt{\Delta^2 +|V|^2}
  \end{equation}
\end{subequations}
where $\Delta \equiv \Delta_0 - \hbar\omega/2$ is the detuning, $\phi = \arg{V}$ is the coupling phase, $\theta \equiv \arccos(\Delta/\Omega) = 
\arcsin(|V|/\Omega)$ is the mixing angle and $\Omega$ is the generalized Rabi frequency.

Consider then   driving this system  by a frequency modulated field with time varying $\omega \to \omega_t$.
The resulting Hamiltonian can be expressed as a modulation of the form discussed above with $\hat{\widetilde{H}}(t) = \hat{H}_0(t; 
\omega_t)$, that sweeps through Hamiltonians in the family described by Eq. (\ref{eq:RabiHam}).
In general, this problem does not admit a closed form solution, but is tractable through direct numerical propagation of the TDSE.
However, in the limit where the modulation changes sufficiently slowly the dynamics can be solved using the generalized adiabatic theorem 
for isolated systems [Eq. (\ref{eq:closedadiab})].

To define the domain in which this theorem applies consider the modulation derivative,
\begin{equation}
  \label{eq:RabiModDeriv}
  \frac{\partial}{\partial \omega} \hat{\widetilde{H}}(t;\omega) =
    \begin{pmatrix}
      0 & itV e^{i\omega t} \\
      -itV^*e^{-i\omega t} & 0
    \end{pmatrix}
\end{equation}
which characterizes the effect of the modulation on the driving field.
Given  Eq. (\ref{eqs:RabiSol}), we have 
\begin{equation}
  \label{eq:RabiDerivEl}
  \ket{-(t;\omega)}  \frac{\partial}{\partial \omega} \hat{\widetilde{H}}(t;\omega) \bra{+(t; \omega)} = i\frac{V\Delta}{\Omega} t,
\end{equation}
required for the generalized adiabatic theorem.

Substituting Eqs. (\ref{eq:RabiDerivEl}) and (\ref{eq:RabiModDeriv}) into Eq. (\ref{eq:closedadiab}) shows that the generalized adiabatic 
theorem applies when
\begin{equation}
  \label{eq:RabiAdiabaticCondition}
  \max_{0\leq t\leq\tau}\left\lbrace \frac{|\dot\omega_t| |V| \Delta_t }{2\Omega_t^2}t\right\rbrace\ll \min_{0\leq t\leq\tau}\lbrace 
  2|\Omega_t|\rbrace,
\end{equation}
where we have indicated the parameters that change over time upon frequency modulation using the subscript $t$.
Here $|\Omega_t|= \sqrt{\Delta_t^2 +|V|^2}$.

\begin{figure}[htbp]
\includegraphics[width=0.8\textwidth]{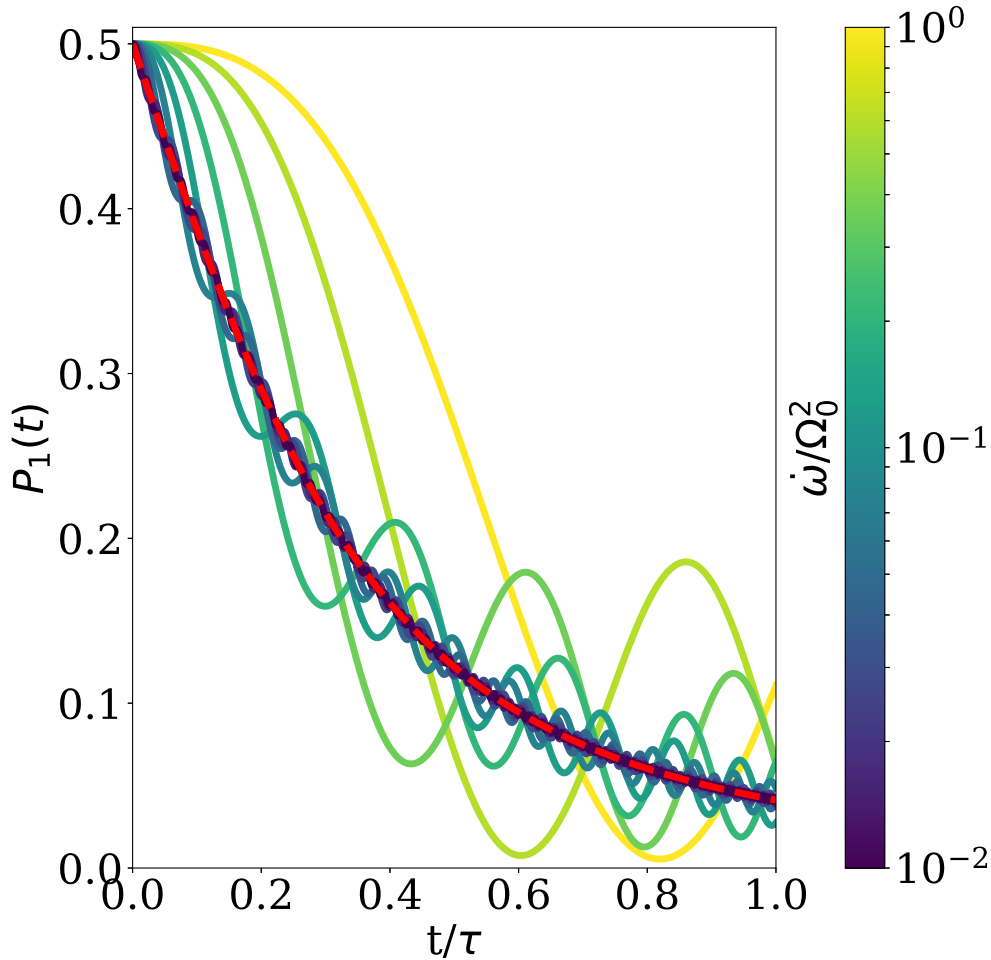}
\caption{Dynamics of a two-level system driven by a frequency modulated field. The frequency of the driving field is swept linearly from a 
resonant frequency $\hbar\omega_0 = \Delta_0$ to $\hbar\omega_\tau = 5\Delta_0$ with sweep velocity $\dot\omega$ (i.e. $\omega_t = \Delta_0 +\dot\omega t$). This transformation takes place over a time interval $\tau = 4\Delta_0/\dot\omega$. The 
yellow to blue traces show the dynamics under a range of sweep velocities while the red dashed line shows the adiabatic prediction expected for 
infinitely slow sweep velocities.}
\label{fig:rabi}
\end{figure}

When Eq. (\ref{eq:RabiAdiabaticCondition}) is well satisfied, then a system initialized in one of the generalized adiabats $\ket\pm$ will remain 
in that state at all times.
To demonstrate  this, we consider  a simple linear modulation $\omega_t = \omega_0 + \dot\omega t$, where $\dot\omega$ is the constant 
frequency sweep velocity, and 
 where the field,  initially in resonance $\Delta_{t=0} = 0$, is swept linearly to $\Delta_{t=\tau} = 4\Delta_0$.
In Fig. \ref{fig:rabi}, numerically exact dynamics are obtained for a range of sweep velocities through Runge-Kutta integration of the TDSE.
These dynamics are then compared to the adiabatic theorem predictions, showing excellent agreement for slow modulations.
(We note that the time taken to perform this modulation varies significantly for different sweep velocities, and we have plotted the dynamics in 
Fig. \ref{fig:rabi} on a normalized time axis $t/\tau$.)

This example is helpful in highlighting the construction and flexibility of the AMT.
The adiabatic speed limit is not set by the energy gap but rather by the Rabi frequency  $\Omega$ on the right hand side of  
Eq. (\ref{eq:RabiAdiabaticCondition}).
This feature, typical of the AMT, is remarkable since it allows for manipulation of the adiabatic time scales
simply by increasing the intensity of the driving field $V$, and hence $|\Omega_t|$, and allows 
for adiabatic transformations of near degenerate systems on manageable timescales.
Moreover, the normal modes, given by the Floquet states $\ket\pm$, follow the dynamics of the system under fixed frequency driving.
By following these oscillatory dynamics the oscillations of the driving field that typically lead to the failure of traditional AT's are removed from consideration
\cite{amin_consistency_2009,du_experimental_2008,tong_quantitative_2005,marzlin_inconsistency_2004}.
In this particularly simple case where only one frequency of light drives the system, this is equivalent to moving to a rotating reference frame using the interaction picture.
The instantaneous normal modes can therefore be thought of as a more general method for following the dynamics of the system when an interacting reference frame cannot be defined.

\subsubsection{Generalizing Adiabatic Experiments}
\label{sec:closed}

As an indication of the additional role of the AMT in isolated systems, note that it allows a wide range of experimental applications beyond that of the traditional AT.
Currently, experimental Adiabatic Rapid Passage (ARP) 
and Stimulated Raman Adiabatic Passage (STIRAP) techniques have combined the rotating wave approximation with a rotating reference frame in order to 
apply the traditional AT to optically driven systems.
  While this approach can be effective for treating systems driven by near-resonant monochromatic lasers or transform limited pulses, it imposes 
  several  restrictions on the types of experiments that can be realized.
  The AMT, however,  removes many of these limitations, allowing for  driving beyond the Rotating Wave Approximation, driving of  a transition by 
  multichromatic fields,  and  driving of more simultaneous transitions than is allowed under  standard conditions.
  The removal of these limitations can have significant effects on  possible experimental techniques.
  (For example, a recently derived adiabaticity condition \cite{wang_necessary_2016}  allowed for the use of specifically designed pulse sequences 
  to realize adiabatic transformations in finite time. \cite{xu_breaking_2019})
  Moreover, the AMT allows for the adiabatic treatment of new phenomena that lie far outside of the rotating wave regime, such as Sisyphus 
  cooling, and can address experimental challenges that are intractable using simpler driving protocols, such  implementing 
  Stimulated Hyper-Raman Adiabatic Passage in the presence of Autler-Townes shifts from spectator states 
  \cite{guerin_stimulated_1998,yatsenko_stimulated_1998}.

To appreciate the advantages afforded by the AMT, consider issues in the application of the traditional AT to oscillatory Hamiltonians.  This 
has been carried out in the Interaction Picture by moving to a rotating reference frame where the oscillatory Hamiltonian time-dependence can be removed.
  If this can not be done, then the Marzlin-Sanders inconsistency \cite{marzlin_inconsistency_2004} prevents  application of the traditional 
  adiabatic theorem.
Hence, applications of the traditional AT are limited to systems where an appropriate choice of basis and reference 
  Hamiltonian can be found to remove all oscillatory time-dependence.

  Consider then the limits of the transformations possible by an appropriate choice of reference Hamiltonian in the Interaction picture. Let
  $H(t)$ be an arbitrary time-dependent Hamiltonian
  \begin{equation}
    \label{eq:RWA}
    H(t) = \sum_{i} E_i(t)\ket{i}\bra{i} + \sum_{ij}V_{ij}(t) \ket{i}\bra{j},
  \end{equation}
  where $E_i(t)$ are the time-dependent diagonal energies and $V_{ij}(t)$ the off-diagonal couplings.

An interaction picture, such as the commonly used rotating reference frame, is defined by a choice of a reference Hamiltonian $H_0$.
  For simplicity, we can work in the eigenbasis of the reference Hamiltonian to give the representation $H_0 = \sum_i E^{(0)}_i\ket{i}\bra{i}$.
  Equation (\ref{eq:RWA}) can  then be rewritten in the interaction picture to give
  \begin{equation}
    \label{eq:RWA-Ham}
    H_I(t) = e^{i\frac{H_0(t)}{\hbar} t}(H(t)-H_0)e^{-i\frac{H_0(t)}{\hbar} t} = \sum_{i}(E_i(t) - E^{(0)}_i)\ket{i}\bra{i} +\sum_{ij} V_{ij}(t) 
    e^{-i\frac{E^{(0)}_{ij}}{\hbar}t}\ket{i}\bra{j}
  \end{equation}
  where $E^{(0)}_{ij} = E^{(0)}_i - E^{(0)}_j$.

Considering the conditions under which oscillatory time-dependence can be removed from Eq. (\ref{eq:RWA-Ham}) provides a set of conditions for 
when AT based approaches can be applied.
This requires a basis in which the diagonal elements of $H(t)$ have no oscillatory 
time-dependence,
defining the eigenbasis of the reference Hamiltonian $H_0$.
The off diagonal elements show that the only way to eliminate oscillatory time-dependence in $V_{ij}(t)$ is if $V_{ij}\propto e^{-i\omega_{ij} t}$ for a 
frequency $\omega_{ij} = E^{(0)}_{ij}/\hbar$. 
This condition that prescribes a specific functional form for the time-dependence of $V_{ij}$ is very restrictive and suggests that the strategy of removing oscillatory Hamiltonian time-dependence will only be effective in specific situations.
However, this approach has been successful since the oscillatory functional form $V_{ij}\propto e^{-i\omega_{ij} t}$ is precisely what is seen when modelling optically driven transitions in the rotating wave approximation.

This condition, however, does limit the application of the traditional adiabatic theorem to specific types of optical excitation.
First, the excitation must be well modelled using the rotating wave approximation.
If this is not the case, then the off-diagonal matrix elements would be real valued (e.g. of the form $V_{ij} \propto \sin(\omega_{ij}t)$) and the interaction picture transformation would leave a residual oscillatory component (e.g. of the form $e^{2i\omega_{ij}t}$).
Moreover, it precludes the driving of one transition by more than one frequency of light, preventing the use of multi-color excitation.
Finally, for a discrete $N$ dimensional system, the $N(N-1)$ off diagonal elements impose up to $N(N-1)$ conditions $\hbar \omega_{ij} = 
E^{(0)}_{ij}$ conditions on the $N$ diagonal elements of $H_0$. Beyond two-dimensional spin systems, it is not guaranteed that it will be 
possible to simultaneously satisfy all of these conditions.
As a result, removal of oscillatory time-dependence can only be guaranteed when a total of $N$ or fewer transitions are driven by an oscillatory 
external field unless additional resonance conditions are met.  The AMT shares none of these limitations.

   \subsection{Open System Adiabatic Turn-on of Incoherent Light}
\label{sec:ex-open}

  A particularly important case in open system quantum mechanics involves the 
  adiabatic turn-on of incoherent radiation that is incident on a molecule and 
 the role of
quantum coherences in biological processes (e.g photosynthesis and vision).
In particular, oscillatory coherences have been observed experimentally
in the excitation of biological molecules with pulsed lasers 
\cite{engel_evidence_2007,zhang_delocalized_2015,ishizaki_quantum_2010,johnson_local_2015}.
By contrast, we have argued, supported by numerical studies, that natural processes rely upon slowly-turned-on
incoherent light, which leads to steady-state transport \cite{jiang_creation_1991,brumer_shedding_2018,dodin_light-induced_2019} with no oscillating coherences.
As shown below, the application of the open system adiabatic theorem proves that
if a system is driven by very-slowly turned on light, then the only coherences that will be observed are those that survive to the steady state.
In particular,  the pulsed laser generated coherences noted above do not survive and are irrelevant under
natural biological conditions.

Specifically, consider a molecular system initially prepared in the absence of a driving light field.
In this case, the system is initially  in a simple equilibrium steady state (i.e.,  with vanishing dynamical frequency).
As such, before the radiation field is turned on, the system is found in a Jordan block with zero eigenvalue.
At $t=0$ an incoherent light field is turned on on a time scale much slower than the dynamics of the molecule,
a typical circumstance in light induced biophysical processes.
This corresponds to a modulation in the limit of $\dot{\lambda}_t \to 0$, ensuring adiabaticity [see Eq. (\ref{eq:openadibatic}]    of the underlying dynamics.
As a result, at all times, the system is found in a Jordan block with vanishing eigenvalue,
that is, in a steady state.
\textit{This  then indicates that the only coherences observed in the slow turn-on limit are those that survive in the steady-state}, 
such as those seen in previous theoretical studies 
\cite{koyu_steady-state_2020,dodin_light-induced_2019,reppert_equilibrium_2019,axelrod_efficient_2018,axelrod_multiple_2019}.

\begin{figure}[htbp]
\includegraphics[width=0.8\textwidth]{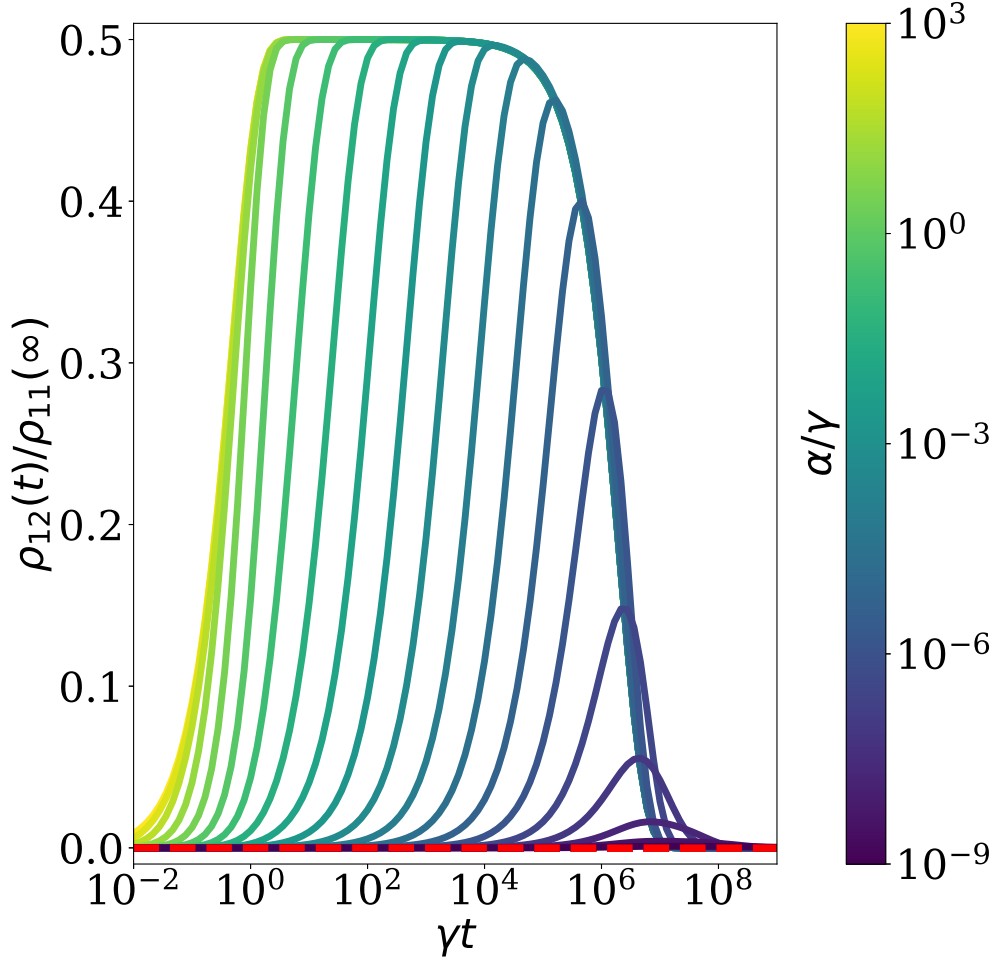}
\caption{Dynamics of a V system driven by slowly turned on incoherent light. The intensity of the exciting field, i.e. the mean photon number, 
turns on with a rate $\alpha$ following an exponential activation function $\overline{n}(t) \propto (1- \exp(-\alpha t))$. We consider a system 
driven by a blackbody source with Temperature $T=5800K$ with ground to excited state excitation energy of $\omega_{ge} = 1.98$ eV. Both excited 
states have a natural line width of $\gamma = 1$ GHz with a splitting of $\Delta = 0.001\hbar\gamma$. The yellow to blue traces show the dynamics 
under a range of turn on rates while the dashed red trace shows the adiabatic turn-on prediction.}
\label{fig:Vsys}
\end{figure}

In many cases, consistency with thermodynamics requires the system to have one steady state given by the Gibbs state $\rho \propto 
\exp(-\hat{H}/k_B T)$ which shows no coherences between non-degenerate energy eigenstates.
This result suggests that the non-steady state coherent dynamics observed under finite turn-on times are a consequence of the rapid turn on of the 
incoherent light field.
Figure \ref{fig:Vsys} discussed below provides a numerical example of
these predictions for a popular generic V-system model \cite{dodin_light-induced_2019,dodin_coherent_2016}.

The ability to show that no coherent dynamics survive the slow turn-on limit of incoherent light without requiring any calculation highlights the 
intuitive power of the adiabatic theorems derived in this letter.
Notably, this situation is far beyond the bounds of applicability of previous adiabaticity condition as the system dynamics are driven by a noisy 
incoherent light field.
This produces a driving Hamiltonian that fluctuates extremely rapidly, on the order of 1 fs, much faster than the energy gap between states and 
consequently the underlying molecular dynamics.
A numerical example follows below.

Consider then the basic minimal model: an open three level system under irradiation by slowly turned-on incoherent light.
This model has been previously treated where numerical results showed \cite{dodin_coherent_2016} that coherent dynamics, i.e. coherences in the 
excited state, vanished under slow turn-on of the exciting field.
The system has a ground energy eigenstate, $\ket{g}$, and two excited states, $\ket{e_1}$ and $\ket{e_2}$ that are separated by an energy $\Delta$ 
and is excited by an incoherent light source with time-dependent intensity.
The dynamics of this system, in the weak coupling limit, is characterized by the following Partial Secular Bloch-Redfield Master equations for the 
matrix elements of the density operator $\hat\rho$:
\begin{subequations}
  \label{eqs:BREOM}
  \begin{equation}
    \label{eq:BRrhogg}
    \dot\rho_{gg} = -(r_1(t)+r_2(t)) \rho_{gg} + (r_1(t) +\gamma_1) \rho_{e_1e_1} + (r_2(t) + \gamma_2)\rho_{e_2e_2} + 2p(\sqrt{r_1(t)r_2(t)} 
    +\sqrt{\gamma_1\gamma_2})\rho_{e_1e_2}^R
  \end{equation}
  \begin{equation}
    \label{eq:BRrhoee}
    \dot\rho_{e_ie_i} = r_i(t)\rho_{gg} - (r_i(t) + \gamma_i)\rho_{e_ie_i} -p(\sqrt{r_1(t)r_2(t)} + \sqrt{\gamma_1\gamma_2})\rho_{e_1e_2}^R
  \end{equation}
  \begin{equation}
    \label{eq:BRrhocoh}
    \dot\rho_{e_1e_2} = -\half(r_1(t) +r_2(t) + \gamma_1 + \gamma_2 + 2i\Delta) \rho_{e_1e_2} + 
    \frac{p}{2}\sqrt{r_1(t)r_2(t)}(2\rho_{gg}-\rho_{e_1e_1}-\rho_{e_2e_2}) - \frac{p}{2}\sqrt{\gamma_1\gamma_2}(\rho_{e_1e_1} + \rho_{e_2e_2})
  \end{equation}
\end{subequations}
where $\gamma_i$ is the spontaneous emission rate and $r_i(t)$ is the rate of excitation to (and stimulated emission from) state $\ket{e_i}$.
The alignment parameter $p$ characterizes the probability of simultaneous excitation to states $\ket{e_1}$ and $\ket{e_2}$ and is defined as
\begin{equation}
  \label{eq:pdef}
  p \equiv \frac{\bm{\mu_1}\cdot\bm{\mu_2}}{\mu_1\mu_2}
\end{equation}
where $\bm{\mu_i}$ is the transition dipole moment between the ground state $\ket{g}$ and excited state $\ket{e_i}$, assumed real for simplicity.
The spontaneous emission and excitation rates are related by detailed balance to give $r_i(t) = \gamma_i \overline{n}(t)$ where $\overline{n}(t)$ 
is the mean occupation number of the resonant thermal field mode.
It is the slow turn-on of the incoherent light that leads to a time dependent occupation number $\overline{n}(t)$ of the thermal field reflecting 
it's time dependent intensity.
That is, the slow modulation of the Liouvillian arises due to slow changes in the statistics of the exciting field, in this case in the mean 
number of photons in the thermal field modes.

These equations of motion can be analytically solved giving several distinct dynamical regimes.
In particular, in the limit where $\overline\gamma/\Delta_P \gg 1$, with $\overline\gamma = (\gamma_1 +\gamma_2)/2$ and $\Delta_p = \sqrt{\Delta^2 
+ (1-p^2)\gamma_1\gamma_2}$, a regime examined below,  the system shows long-lived quasi-stationary coherences that eventually decay to give an incoherent thermal state 
\cite{tscherbul_long-lived_2014}.
Under excitation by a field with time-dependent intensity of the form $\overline{n}(t) = \overline{n}(1 - \exp(-\alpha t))$, the dynamics are 
given by
\begin{subequations}
  \label{eqs:BRDyn}
  \begin{equation}
    \label{eq:BRDynEE}
    \rho_{e_ie_i}(t) = \frac{1}{2\overline{\gamma}}\left\lbrace
      \frac{r_j}{\frac{\Delta_p^2}{2\overline\gamma}-\alpha}\left[ \frac{\Delta_p^2}{2\overline\gamma}\left(1-e^{-\alpha t}\right) - 
      \alpha\left(1-e^{-\frac{\Delta_p^2}{2\overline\gamma}t}\right)\right]
      +\frac{r_i}{2\overline\gamma - \alpha}\left[2\overline\gamma\left(1-e^{-\alpha t}\right) - \alpha \left(1-e^{-2\overline\gamma 
      t}\right)\right]
    \right\rbrace
  \end{equation}
   \begin{equation}
    \label{eq:BRDynCoh}
    \rho_{e_1e_2}(t) = \frac{p\sqrt{r_1r_2}}{2\overline{\gamma}}\left\lbrace
      \frac{1}{\frac{\Delta_p^2}{2\overline\gamma}-\alpha}\left[ \frac{\Delta_p^2}{2\overline\gamma}\left(1-e^{-\alpha t}\right) - 
      \alpha\left(1-e^{-\frac{\Delta_p^2}{2\overline\gamma}t}\right)\right]
      +\frac{1}{2\overline\gamma - \alpha}\left[2\overline\gamma\left(1-e^{-\alpha t}\right) - \alpha \left(1-e^{-2\overline\gamma 
      t}\right)\right]
    \right\rbrace
  \end{equation}
\end{subequations}
with $\overline\gamma$ and $\Delta_p$ defined above.
A detailed derivation of Eq. (\ref{eqs:BRDyn}) along with an in depth discussion of the underlying physics are given in Ref. 
\cite{dodin_coherent_2016}.

While Eq. (\ref{eqs:BREOM}) is complicated, the dynamics are qualitatively simple.
When the turn on time $\tau_r = \alpha^{-1}$ is shorter than the coherence lifetime $2\overline\gamma/\Delta_p^2$, the quasi-stationary coherences 
lead to a potentially long-lived superposition of excited states which rises on a timescale $\tau_r$.
The coherences then decay on a timescale $2\overline \gamma/\Delta_p^2$.
As the turn on time becomes comparable to or longer than the coherence lifetime, the quasi-stationary superposition is not fully excited, leading 
to a decrease in the generated coherences that disappear entirely in the $\tau_r\to\infty$ limit.
These dynamics are shown for a variety of turn on rates, $\alpha$, in the solid traces of Fig. \ref{fig:Vsys} for the case of $p=1$, $\gamma_1 = 
\gamma_2$.

These traces can be compared to the dashed red trace which shows the predictions of the open system adiabatic theorem.
As discussed above, since the system is initialized in a steady state of the zero field case, a sufficiently slow turn on leads to dynamics that 
remain in the instantaneous steady state at $\overline{n}(t)$.
In this example involving only a single bath, the steady state is the thermal Gibbs state with a temperature determined by $\overline{n}(t)$,  and therefore 
shows no coherences.
Consequently, as illustrated in this example, the open systems adiabatic theorem guarantees that a sufficiently slow turn-on of an incoherent 
field will only show coherences that survive in the steady-state.
In the case of two baths, steady state coherences associated with transport will
persist \cite{dodin_light-induced_2019,koyu_steady-state_2020}.

\section{Summary}

We have presented a new set of generalized adiabatic theorems that apply to rapidly fluctuating fields with slowly varying 
properties.
These results express the dynamics of the modulated field in terms of the dynamical normal modes of time dependent reference Hamiltonians.
The resulting adiabaticity conditions take a form that is conceptually similar to preexisting theorems but bounds adiabaticity based on the rate 
of a slowly changing transformation rather than the rapid underlying field. 
These results significantly extend the applicability of the adiabatic theorem and its underlying intuition to, e.g., the domain of optical 
excitations that play a central role in atomic, molecular and optical physics.

{\bf Acknowledgments} This work was supported by the US Air Force Office of Scientific
Research Under Contract Number FA9550-17-1-0310 and FA9550-20-1-0354.

\bibliography{adiabatic} % Omit .bib extension

\begin{thebibliography}{46}%
\makeatletter
\providecommand \@ifxundefined [1]{%
 \@ifx{#1\undefined}
}%
\providecommand \@ifnum [1]{%
 \ifnum #1\expandafter \@firstoftwo
 \else \expandafter \@secondoftwo
 \fi
}%
\providecommand \@ifx [1]{%
 \ifx #1\expandafter \@firstoftwo
 \else \expandafter \@secondoftwo
 \fi
}%
\providecommand \natexlab [1]{#1}%
\providecommand \enquote  [1]{``#1''}%
\providecommand \bibnamefont  [1]{#1}%
\providecommand \bibfnamefont [1]{#1}%
\providecommand \citenamefont [1]{#1}%
\providecommand \href@noop [0]{\@secondoftwo}%
\providecommand \href [0]{\begingroup \@sanitize@url \@href}%
\providecommand \@href[1]{\@@startlink{#1}\@@href}%
\providecommand \@@href[1]{\endgroup#1\@@endlink}%
\providecommand \@sanitize@url [0]{\catcode `\\12\catcode `\$12\catcode
  `\&12\catcode `\#12\catcode `\^12\catcode `\_12\catcode `\%12\relax}%
\providecommand \@@startlink[1]{}%
\providecommand \@@endlink[0]{}%
\providecommand \url  [0]{\begingroup\@sanitize@url \@url }%
\providecommand \@url [1]{\endgroup\@href {#1}{\urlprefix }}%
\providecommand \urlprefix  [0]{URL }%
\providecommand \Eprint [0]{\href }%
\providecommand \doibase [0]{http://dx.doi.org/}%
\providecommand \selectlanguage [0]{\@gobble}%
\providecommand \bibinfo  [0]{\@secondoftwo}%
\providecommand \bibfield  [0]{\@secondoftwo}%
\providecommand \translation [1]{[#1]}%
\providecommand \BibitemOpen [0]{}%
\providecommand \bibitemStop [0]{}%
\providecommand \bibitemNoStop [0]{.\EOS\space}%
\providecommand \EOS [0]{\spacefactor3000\relax}%
\providecommand \BibitemShut  [1]{\csname bibitem#1\endcsname}%
\let\auto@bib@innerbib\@empty
%</preamble>
\bibitem [{\citenamefont {Born}\ and\ \citenamefont
  {Fock}(1928)}]{born_beweis_1928}%
  \BibitemOpen
  \bibfield  {author} {\bibinfo {author} {\bibfnamefont {M.}~\bibnamefont
  {Born}}\ and\ \bibinfo {author} {\bibfnamefont {V.}~\bibnamefont {Fock}},\
  }\href {\doibase 10.1007/BF01343193} {\bibfield  {journal} {\bibinfo
  {journal} {Zeitschrift fur Physik}\ }\textbf {\bibinfo {volume} {51}},\
  \bibinfo {pages} {165} (\bibinfo {year} {1928})}\BibitemShut {NoStop}%
\bibitem [{\citenamefont {Kato}(1950)}]{kato_adiabatic_1950}%
  \BibitemOpen
  \bibfield  {author} {\bibinfo {author} {\bibfnamefont {T.}~\bibnamefont
  {Kato}},\ }\href {\doibase 10.1143/JPSJ.5.435} {\bibfield  {journal}
  {\bibinfo  {journal} {Journal of the Physical Society of Japan}\ }\textbf
  {\bibinfo {volume} {5}},\ \bibinfo {pages} {435} (\bibinfo {year}
  {1950})}\BibitemShut {NoStop}%
\bibitem [{\citenamefont {Sarandy}\ and\ \citenamefont
  {Lidar}(2005)}]{sarandy_adiabatic_2005}%
  \BibitemOpen
  \bibfield  {author} {\bibinfo {author} {\bibfnamefont {M.~S.}\ \bibnamefont
  {Sarandy}}\ and\ \bibinfo {author} {\bibfnamefont {D.~A.}\ \bibnamefont
  {Lidar}},\ }\href {\doibase 10.1103/PhysRevA.71.012331} {\bibfield  {journal}
  {\bibinfo  {journal} {Physical Review A}\ }\textbf {\bibinfo {volume} {71}},\
  \bibinfo {pages} {012331} (\bibinfo {year} {2005})}\BibitemShut {NoStop}%
\bibitem [{\citenamefont {Messiah}(2017)}]{messiah_quantum_2017}%
  \BibitemOpen
  \bibfield  {author} {\bibinfo {author} {\bibfnamefont {A.}~\bibnamefont
  {Messiah}},\ }\href@noop {} {\emph {\bibinfo {title} {Quantum {Mechanics}:
  {Two} {Volumes} {Bound} {As} {One}}}}\ (\bibinfo  {publisher} {Dover
  Publications},\ \bibinfo {address} {Mineola, N.Y},\ \bibinfo {year}
  {2017})\BibitemShut {NoStop}%
\bibitem [{\citenamefont {Landau}(1932)}]{landau_zur_1932}%
  \BibitemOpen
  \bibfield  {author} {\bibinfo {author} {\bibfnamefont {L.~D.}\ \bibnamefont
  {Landau}},\ }\href@noop {} {\bibfield  {journal} {\bibinfo  {journal} {Phys.
  Z. Sowjetunion}\ }\textbf {\bibinfo {volume} {2}},\ \bibinfo {pages} {1}
  (\bibinfo {year} {1932})}\BibitemShut {NoStop}%
\bibitem [{\citenamefont {Zener}(1932)}]{zener_non-adiabatic_1932}%
  \BibitemOpen
  \bibfield  {author} {\bibinfo {author} {\bibfnamefont {C.}~\bibnamefont
  {Zener}},\ }\href {\doibase 10.1098/rspa.1932.0165} {\bibfield  {journal}
  {\bibinfo  {journal} {Proc. R. Soc. Lond. A}\ }\textbf {\bibinfo {volume}
  {137}},\ \bibinfo {pages} {696} (\bibinfo {year} {1932})}\BibitemShut
  {NoStop}%
\bibitem [{\citenamefont {Albash}\ and\ \citenamefont
  {Lidar}(2018)}]{albash_adiabatic_2018}%
  \BibitemOpen
  \bibfield  {author} {\bibinfo {author} {\bibfnamefont {T.}~\bibnamefont
  {Albash}}\ and\ \bibinfo {author} {\bibfnamefont {D.~A.}\ \bibnamefont
  {Lidar}},\ }\href {\doibase 10.1103/RevModPhys.90.015002} {\bibfield
  {journal} {\bibinfo  {journal} {Reviews of Modern Physics}\ }\textbf
  {\bibinfo {volume} {90}},\ \bibinfo {pages} {015002} (\bibinfo {year}
  {2018})}\BibitemShut {NoStop}%
\bibitem [{\citenamefont {Barends}\ \emph {et~al.}(2016)\citenamefont
  {Barends}, \citenamefont {Shabani}, \citenamefont {Lamata}, \citenamefont
  {Kelly}, \citenamefont {Mezzacapo}, \citenamefont {Heras}, \citenamefont
  {Babbush}, \citenamefont {Fowler}, \citenamefont {Campbell}, \citenamefont
  {Chen}, \citenamefont {Chen}, \citenamefont {Chiaro}, \citenamefont
  {Dunsworth}, \citenamefont {Jeffrey}, \citenamefont {Lucero}, \citenamefont
  {Megrant}, \citenamefont {Mutus}, \citenamefont {Neeley}, \citenamefont
  {Neill}, \citenamefont {O’Malley}, \citenamefont {Quintana}, \citenamefont
  {Roushan}, \citenamefont {Sank}, \citenamefont {Vainsencher}, \citenamefont
  {Wenner}, \citenamefont {White}, \citenamefont {Solano}, \citenamefont
  {Neven},\ and\ \citenamefont {Martinis}}]{barends_digitized_2016}%
  \BibitemOpen
  \bibfield  {author} {\bibinfo {author} {\bibfnamefont {R.}~\bibnamefont
  {Barends}}, \bibinfo {author} {\bibfnamefont {A.}~\bibnamefont {Shabani}},
  \bibinfo {author} {\bibfnamefont {L.}~\bibnamefont {Lamata}}, \bibinfo
  {author} {\bibfnamefont {J.}~\bibnamefont {Kelly}}, \bibinfo {author}
  {\bibfnamefont {A.}~\bibnamefont {Mezzacapo}}, \bibinfo {author}
  {\bibfnamefont {U.~L.}\ \bibnamefont {Heras}}, \bibinfo {author}
  {\bibfnamefont {R.}~\bibnamefont {Babbush}}, \bibinfo {author} {\bibfnamefont
  {A.~G.}\ \bibnamefont {Fowler}}, \bibinfo {author} {\bibfnamefont
  {B.}~\bibnamefont {Campbell}}, \bibinfo {author} {\bibfnamefont
  {Y.}~\bibnamefont {Chen}}, \bibinfo {author} {\bibfnamefont {Z.}~\bibnamefont
  {Chen}}, \bibinfo {author} {\bibfnamefont {B.}~\bibnamefont {Chiaro}},
  \bibinfo {author} {\bibfnamefont {A.}~\bibnamefont {Dunsworth}}, \bibinfo
  {author} {\bibfnamefont {E.}~\bibnamefont {Jeffrey}}, \bibinfo {author}
  {\bibfnamefont {E.}~\bibnamefont {Lucero}}, \bibinfo {author} {\bibfnamefont
  {A.}~\bibnamefont {Megrant}}, \bibinfo {author} {\bibfnamefont {J.~Y.}\
  \bibnamefont {Mutus}}, \bibinfo {author} {\bibfnamefont {M.}~\bibnamefont
  {Neeley}}, \bibinfo {author} {\bibfnamefont {C.}~\bibnamefont {Neill}},
  \bibinfo {author} {\bibfnamefont {P.~J.~J.}\ \bibnamefont {O’Malley}},
  \bibinfo {author} {\bibfnamefont {C.}~\bibnamefont {Quintana}}, \bibinfo
  {author} {\bibfnamefont {P.}~\bibnamefont {Roushan}}, \bibinfo {author}
  {\bibfnamefont {D.}~\bibnamefont {Sank}}, \bibinfo {author} {\bibfnamefont
  {A.}~\bibnamefont {Vainsencher}}, \bibinfo {author} {\bibfnamefont
  {J.}~\bibnamefont {Wenner}}, \bibinfo {author} {\bibfnamefont {T.~C.}\
  \bibnamefont {White}}, \bibinfo {author} {\bibfnamefont {E.}~\bibnamefont
  {Solano}}, \bibinfo {author} {\bibfnamefont {H.}~\bibnamefont {Neven}}, \
  and\ \bibinfo {author} {\bibfnamefont {J.~M.}\ \bibnamefont {Martinis}},\
  }\href {\doibase 10.1038/nature17658} {\bibfield  {journal} {\bibinfo
  {journal} {Nature}\ }\textbf {\bibinfo {volume} {534}},\ \bibinfo {pages}
  {222} (\bibinfo {year} {2016})}\BibitemShut {NoStop}%
\bibitem [{\citenamefont {Torrontegui}\ \emph {et~al.}(2017)\citenamefont
  {Torrontegui}, \citenamefont {Lizuain}, \citenamefont {González-Resines},
  \citenamefont {Tobalina}, \citenamefont {Ruschhaupt}, \citenamefont
  {Kosloff},\ and\ \citenamefont {Muga}}]{torrontegui_energy_2017}%
  \BibitemOpen
  \bibfield  {author} {\bibinfo {author} {\bibfnamefont {E.}~\bibnamefont
  {Torrontegui}}, \bibinfo {author} {\bibfnamefont {I.}~\bibnamefont
  {Lizuain}}, \bibinfo {author} {\bibfnamefont {S.}~\bibnamefont
  {González-Resines}}, \bibinfo {author} {\bibfnamefont {A.}~\bibnamefont
  {Tobalina}}, \bibinfo {author} {\bibfnamefont {A.}~\bibnamefont
  {Ruschhaupt}}, \bibinfo {author} {\bibfnamefont {R.}~\bibnamefont {Kosloff}},
  \ and\ \bibinfo {author} {\bibfnamefont {J.~G.}\ \bibnamefont {Muga}},\
  }\href {\doibase 10.1103/PhysRevA.96.022133} {\bibfield  {journal} {\bibinfo
  {journal} {Physical Review A}\ }\textbf {\bibinfo {volume} {96}},\ \bibinfo
  {pages} {022133} (\bibinfo {year} {2017})}\BibitemShut {NoStop}%
\bibitem [{\citenamefont {del Campo}(2013)}]{del_campo_shortcuts_2013}%
  \BibitemOpen
  \bibfield  {author} {\bibinfo {author} {\bibfnamefont {A.}~\bibnamefont {del
  Campo}},\ }\href {\doibase 10.1103/PhysRevLett.111.100502} {\bibfield
  {journal} {\bibinfo  {journal} {Physical Review Letters}\ }\textbf {\bibinfo
  {volume} {111}},\ \bibinfo {pages} {100502} (\bibinfo {year}
  {2013})}\BibitemShut {NoStop}%
\bibitem [{\citenamefont {Torrontegui}\ \emph {et~al.}(2013)\citenamefont
  {Torrontegui}, \citenamefont {Ibáñez}, \citenamefont {Martínez-Garaot},
  \citenamefont {Modugno}, \citenamefont {del Campo}, \citenamefont
  {Guéry-Odelin}, \citenamefont {Ruschhaupt}, \citenamefont {Chen},\ and\
  \citenamefont {Muga}}]{torrontegui_chapter_2013}%
  \BibitemOpen
  \bibfield  {author} {\bibinfo {author} {\bibfnamefont {E.}~\bibnamefont
  {Torrontegui}}, \bibinfo {author} {\bibfnamefont {S.}~\bibnamefont
  {Ibáñez}}, \bibinfo {author} {\bibfnamefont {S.}~\bibnamefont
  {Martínez-Garaot}}, \bibinfo {author} {\bibfnamefont {M.}~\bibnamefont
  {Modugno}}, \bibinfo {author} {\bibfnamefont {A.}~\bibnamefont {del Campo}},
  \bibinfo {author} {\bibfnamefont {D.}~\bibnamefont {Guéry-Odelin}}, \bibinfo
  {author} {\bibfnamefont {A.}~\bibnamefont {Ruschhaupt}}, \bibinfo {author}
  {\bibfnamefont {X.}~\bibnamefont {Chen}}, \ and\ \bibinfo {author}
  {\bibfnamefont {J.~G.}\ \bibnamefont {Muga}},\ }in\ \href {\doibase
  10.1016/B978-0-12-408090-4.00002-5} {\emph {\bibinfo {booktitle} {Advances
  {In} {Atomic}, {Molecular}, and {Optical} {Physics}}}},\ \bibinfo {series}
  {Advances in {Atomic}, {Molecular}, and {Optical} {Physics}}, Vol.~\bibinfo
  {volume} {62},\ \bibinfo {editor} {edited by\ \bibinfo {editor}
  {\bibfnamefont {E.}~\bibnamefont {Arimondo}}, \bibinfo {editor}
  {\bibfnamefont {P.~R.}\ \bibnamefont {Berman}}, \ and\ \bibinfo {editor}
  {\bibfnamefont {C.~C.}\ \bibnamefont {Lin}}}\ (\bibinfo  {publisher}
  {Academic Press},\ \bibinfo {year} {2013})\ pp.\ \bibinfo {pages}
  {117--169}\BibitemShut {NoStop}%
\bibitem [{\citenamefont {Campo}\ and\ \citenamefont
  {Boshier}(2012)}]{campo_shortcuts_2012}%
  \BibitemOpen
  \bibfield  {author} {\bibinfo {author} {\bibfnamefont {A.~d.}\ \bibnamefont
  {Campo}}\ and\ \bibinfo {author} {\bibfnamefont {M.~G.}\ \bibnamefont
  {Boshier}},\ }\href {\doibase 10.1038/srep00648} {\bibfield  {journal}
  {\bibinfo  {journal} {Scientific Reports}\ }\textbf {\bibinfo {volume} {2}},\
  \bibinfo {pages} {648} (\bibinfo {year} {2012})}\BibitemShut {NoStop}%
\bibitem [{\citenamefont {Amin}(2009)}]{amin_consistency_2009}%
  \BibitemOpen
  \bibfield  {author} {\bibinfo {author} {\bibfnamefont {M.~H.~S.}\
  \bibnamefont {Amin}},\ }\href {\doibase 10.1103/PhysRevLett.102.220401}
  {\bibfield  {journal} {\bibinfo  {journal} {Physical Review Letters}\
  }\textbf {\bibinfo {volume} {102}},\ \bibinfo {pages} {220401} (\bibinfo
  {year} {2009})},\ \bibinfo {note} {publisher: American Physical
  Society}\BibitemShut {NoStop}%
\bibitem [{\citenamefont {Du}\ \emph {et~al.}(2008)\citenamefont {Du},
  \citenamefont {Hu}, \citenamefont {Wang}, \citenamefont {Wu}, \citenamefont
  {Zhao},\ and\ \citenamefont {Suter}}]{du_experimental_2008}%
  \BibitemOpen
  \bibfield  {author} {\bibinfo {author} {\bibfnamefont {J.}~\bibnamefont
  {Du}}, \bibinfo {author} {\bibfnamefont {L.}~\bibnamefont {Hu}}, \bibinfo
  {author} {\bibfnamefont {Y.}~\bibnamefont {Wang}}, \bibinfo {author}
  {\bibfnamefont {J.}~\bibnamefont {Wu}}, \bibinfo {author} {\bibfnamefont
  {M.}~\bibnamefont {Zhao}}, \ and\ \bibinfo {author} {\bibfnamefont
  {D.}~\bibnamefont {Suter}},\ }\href {\doibase 10.1103/PhysRevLett.101.060403}
  {\bibfield  {journal} {\bibinfo  {journal} {Physical Review Letters}\
  }\textbf {\bibinfo {volume} {101}},\ \bibinfo {pages} {060403} (\bibinfo
  {year} {2008})},\ \bibinfo {note} {publisher: American Physical
  Society}\BibitemShut {NoStop}%
\bibitem [{\citenamefont {Tong}\ \emph {et~al.}(2005)\citenamefont {Tong},
  \citenamefont {Singh}, \citenamefont {Kwek},\ and\ \citenamefont
  {Oh}}]{tong_quantitative_2005}%
  \BibitemOpen
  \bibfield  {author} {\bibinfo {author} {\bibfnamefont {D.~M.}\ \bibnamefont
  {Tong}}, \bibinfo {author} {\bibfnamefont {K.}~\bibnamefont {Singh}},
  \bibinfo {author} {\bibfnamefont {L.~C.}\ \bibnamefont {Kwek}}, \ and\
  \bibinfo {author} {\bibfnamefont {C.~H.}\ \bibnamefont {Oh}},\ }\href
  {\doibase 10.1103/PhysRevLett.95.110407} {\bibfield  {journal} {\bibinfo
  {journal} {Physical Review Letters}\ }\textbf {\bibinfo {volume} {95}},\
  \bibinfo {pages} {110407} (\bibinfo {year} {2005})},\ \bibinfo {note}
  {publisher: American Physical Society}\BibitemShut {NoStop}%
\bibitem [{\citenamefont {Marzlin}\ and\ \citenamefont
  {Sanders}(2004)}]{marzlin_inconsistency_2004}%
  \BibitemOpen
  \bibfield  {author} {\bibinfo {author} {\bibfnamefont {K.-P.}\ \bibnamefont
  {Marzlin}}\ and\ \bibinfo {author} {\bibfnamefont {B.~C.}\ \bibnamefont
  {Sanders}},\ }\href {\doibase 10.1103/PhysRevLett.93.160408} {\bibfield
  {journal} {\bibinfo  {journal} {Physical Review Letters}\ }\textbf {\bibinfo
  {volume} {93}},\ \bibinfo {pages} {160408} (\bibinfo {year} {2004})},\
  \bibinfo {note} {publisher: American Physical Society}\BibitemShut {NoStop}%
\bibitem [{\citenamefont {Malinovsky}\ and\ \citenamefont
  {Krause}(2001)}]{malinovsky_general_2001}%
  \BibitemOpen
  \bibfield  {author} {\bibinfo {author} {\bibfnamefont {V.~S.}\ \bibnamefont
  {Malinovsky}}\ and\ \bibinfo {author} {\bibfnamefont {J.~L.}\ \bibnamefont
  {Krause}},\ }\href {\doibase 10.1007/s100530170212} {\bibfield  {journal}
  {\bibinfo  {journal} {European Physical Journal D}\ }\textbf {\bibinfo
  {volume} {14}},\ \bibinfo {pages} {147} (\bibinfo {year} {2001})}\BibitemShut
  {NoStop}%
\bibitem [{\citenamefont {Vitanov}\ \emph {et~al.}(2017)\citenamefont
  {Vitanov}, \citenamefont {Rangelov}, \citenamefont {Shore},\ and\
  \citenamefont {Bergmann}}]{vitanov_stimulated_2017}%
  \BibitemOpen
  \bibfield  {author} {\bibinfo {author} {\bibfnamefont {N.~V.}\ \bibnamefont
  {Vitanov}}, \bibinfo {author} {\bibfnamefont {A.~A.}\ \bibnamefont
  {Rangelov}}, \bibinfo {author} {\bibfnamefont {B.~W.}\ \bibnamefont {Shore}},
  \ and\ \bibinfo {author} {\bibfnamefont {K.}~\bibnamefont {Bergmann}},\
  }\href {\doibase 10.1103/RevModPhys.89.015006} {\bibfield  {journal}
  {\bibinfo  {journal} {Reviews of Modern Physics}\ }\textbf {\bibinfo {volume}
  {89}},\ \bibinfo {pages} {015006} (\bibinfo {year} {2017})},\ \bibinfo {note}
  {publisher: American Physical Society}\BibitemShut {NoStop}%
\bibitem [{\citenamefont {Bergmann}\ \emph {et~al.}(2015)\citenamefont
  {Bergmann}, \citenamefont {Vitanov},\ and\ \citenamefont
  {Shore}}]{bergmann_perspective_2015}%
  \BibitemOpen
  \bibfield  {author} {\bibinfo {author} {\bibfnamefont {K.}~\bibnamefont
  {Bergmann}}, \bibinfo {author} {\bibfnamefont {N.~V.}\ \bibnamefont
  {Vitanov}}, \ and\ \bibinfo {author} {\bibfnamefont {B.~W.}\ \bibnamefont
  {Shore}},\ }\href {\doibase 10.1063/1.4916903} {\bibfield  {journal}
  {\bibinfo  {journal} {The Journal of Chemical Physics}\ }\textbf {\bibinfo
  {volume} {142}},\ \bibinfo {pages} {170901} (\bibinfo {year} {2015})},\
  \bibinfo {note} {publisher: American Institute of Physics}\BibitemShut
  {NoStop}%
\bibitem [{\citenamefont {Kaufmann}\ \emph {et~al.}(2001)\citenamefont
  {Kaufmann}, \citenamefont {Ekers}, \citenamefont {Gebauer-Rochholz},
  \citenamefont {Mettendorf}, \citenamefont {Keil},\ and\ \citenamefont
  {Bergmann}}]{kaufmann_dissociative_2001}%
  \BibitemOpen
  \bibfield  {author} {\bibinfo {author} {\bibfnamefont {O.}~\bibnamefont
  {Kaufmann}}, \bibinfo {author} {\bibfnamefont {A.}~\bibnamefont {Ekers}},
  \bibinfo {author} {\bibfnamefont {C.}~\bibnamefont {Gebauer-Rochholz}},
  \bibinfo {author} {\bibfnamefont {K.~U.}\ \bibnamefont {Mettendorf}},
  \bibinfo {author} {\bibfnamefont {M.}~\bibnamefont {Keil}}, \ and\ \bibinfo
  {author} {\bibfnamefont {K.}~\bibnamefont {Bergmann}},\ }\href {\doibase
  10.1016/S1387-3806(00)00290-6} {\bibfield  {journal} {\bibinfo  {journal}
  {International Journal of Mass Spectrometry}\ }\bibinfo {series} {Low
  {Energy} {Electron}-{Molecule} {Interactions} ({Stamatovic} honor)},\ \textbf
  {\bibinfo {volume} {205}},\ \bibinfo {pages} {233} (\bibinfo {year}
  {2001})}\BibitemShut {NoStop}%
\bibitem [{\citenamefont {Külz}\ \emph {et~al.}(1996)\citenamefont {Külz},
  \citenamefont {Keil}, \citenamefont {Kortyna}, \citenamefont
  {Schellhaa{\textbackslash}S}, \citenamefont {Hauck}, \citenamefont
  {Bergmann}, \citenamefont {Meyer},\ and\ \citenamefont
  {Weyh}}]{kulz_dissociative_1996}%
  \BibitemOpen
  \bibfield  {author} {\bibinfo {author} {\bibfnamefont {M.}~\bibnamefont
  {Külz}}, \bibinfo {author} {\bibfnamefont {M.}~\bibnamefont {Keil}},
  \bibinfo {author} {\bibfnamefont {A.}~\bibnamefont {Kortyna}}, \bibinfo
  {author} {\bibfnamefont {B.}~\bibnamefont {Schellhaa{\textbackslash}S}},
  \bibinfo {author} {\bibfnamefont {J.}~\bibnamefont {Hauck}}, \bibinfo
  {author} {\bibfnamefont {K.}~\bibnamefont {Bergmann}}, \bibinfo {author}
  {\bibfnamefont {W.}~\bibnamefont {Meyer}}, \ and\ \bibinfo {author}
  {\bibfnamefont {D.}~\bibnamefont {Weyh}},\ }\href {\doibase
  10.1103/PhysRevA.53.3324} {\bibfield  {journal} {\bibinfo  {journal}
  {Physical Review A}\ }\textbf {\bibinfo {volume} {53}},\ \bibinfo {pages}
  {3324} (\bibinfo {year} {1996})},\ \bibinfo {note} {publisher: American
  Physical Society}\BibitemShut {NoStop}%
\bibitem [{\citenamefont {Aikawa}\ \emph {et~al.}(2010)\citenamefont {Aikawa},
  \citenamefont {Akamatsu}, \citenamefont {Hayashi}, \citenamefont {Oasa},
  \citenamefont {Kobayashi}, \citenamefont {Naidon}, \citenamefont {Kishimoto},
  \citenamefont {Ueda},\ and\ \citenamefont {Inouye}}]{aikawa_coherent_2010}%
  \BibitemOpen
  \bibfield  {author} {\bibinfo {author} {\bibfnamefont {K.}~\bibnamefont
  {Aikawa}}, \bibinfo {author} {\bibfnamefont {D.}~\bibnamefont {Akamatsu}},
  \bibinfo {author} {\bibfnamefont {M.}~\bibnamefont {Hayashi}}, \bibinfo
  {author} {\bibfnamefont {K.}~\bibnamefont {Oasa}}, \bibinfo {author}
  {\bibfnamefont {J.}~\bibnamefont {Kobayashi}}, \bibinfo {author}
  {\bibfnamefont {P.}~\bibnamefont {Naidon}}, \bibinfo {author} {\bibfnamefont
  {T.}~\bibnamefont {Kishimoto}}, \bibinfo {author} {\bibfnamefont
  {M.}~\bibnamefont {Ueda}}, \ and\ \bibinfo {author} {\bibfnamefont
  {S.}~\bibnamefont {Inouye}},\ }\href {\doibase
  10.1103/PhysRevLett.105.203001} {\bibfield  {journal} {\bibinfo  {journal}
  {Physical Review Letters}\ }\textbf {\bibinfo {volume} {105}},\ \bibinfo
  {pages} {203001} (\bibinfo {year} {2010})},\ \bibinfo {note} {publisher:
  American Physical Society}\BibitemShut {NoStop}%
\bibitem [{\citenamefont {Dodin}\ and\ \citenamefont
  {Brumer}(2019)}]{dodin_light-induced_2019}%
  \BibitemOpen
  \bibfield  {author} {\bibinfo {author} {\bibfnamefont {A.}~\bibnamefont
  {Dodin}}\ and\ \bibinfo {author} {\bibfnamefont {P.}~\bibnamefont {Brumer}},\
  }\href {\doibase 10.1063/1.5092981} {\bibfield  {journal} {\bibinfo
  {journal} {The Journal of Chemical Physics}\ }\textbf {\bibinfo {volume}
  {150}},\ \bibinfo {pages} {184304} (\bibinfo {year} {2019})}\BibitemShut
  {NoStop}%
\bibitem [{\citenamefont {Dodin}\ \emph {et~al.}(2016)\citenamefont {Dodin},
  \citenamefont {Tscherbul},\ and\ \citenamefont
  {Brumer}}]{dodin_coherent_2016}%
  \BibitemOpen
  \bibfield  {author} {\bibinfo {author} {\bibfnamefont {A.}~\bibnamefont
  {Dodin}}, \bibinfo {author} {\bibfnamefont {T.~V.}\ \bibnamefont
  {Tscherbul}}, \ and\ \bibinfo {author} {\bibfnamefont {P.}~\bibnamefont
  {Brumer}},\ }\href {\doibase 10.1063/1.4972140} {\bibfield  {journal}
  {\bibinfo  {journal} {The Journal of Chemical Physics}\ }\textbf {\bibinfo
  {volume} {145}},\ \bibinfo {pages} {244313} (\bibinfo {year}
  {2016})}\BibitemShut {NoStop}%
\bibitem [{\citenamefont {Cheng}\ and\ \citenamefont
  {Fleming}(2009)}]{cheng_dynamics_2009}%
  \BibitemOpen
  \bibfield  {author} {\bibinfo {author} {\bibfnamefont {Y.-C.}\ \bibnamefont
  {Cheng}}\ and\ \bibinfo {author} {\bibfnamefont {G.~R.}\ \bibnamefont
  {Fleming}},\ }\href {\doibase 10.1146/annurev.physchem.040808.090259}
  {\bibfield  {journal} {\bibinfo  {journal} {Annual Review of Physical
  Chemistry}\ }\textbf {\bibinfo {volume} {60}},\ \bibinfo {pages} {241}
  (\bibinfo {year} {2009})}\BibitemShut {NoStop}%
\bibitem [{\citenamefont {Chenu}\ and\ \citenamefont
  {Scholes}(2015)}]{chenu_coherence_2015}%
  \BibitemOpen
  \bibfield  {author} {\bibinfo {author} {\bibfnamefont {A.}~\bibnamefont
  {Chenu}}\ and\ \bibinfo {author} {\bibfnamefont {G.~D.}\ \bibnamefont
  {Scholes}},\ }\href {\doibase 10.1146/annurev-physchem-040214-121713}
  {\bibfield  {journal} {\bibinfo  {journal} {Annual Review of Physical
  Chemistry}\ }\textbf {\bibinfo {volume} {66}},\ \bibinfo {pages} {69}
  (\bibinfo {year} {2015})}\BibitemShut {NoStop}%
\bibitem [{\citenamefont {Byrne}(1993)}]{byrne_signal_1993}%
  \BibitemOpen
  \bibfield  {author} {\bibinfo {author} {\bibfnamefont {C.~L.}\ \bibnamefont
  {Byrne}},\ }\href@noop {} {\emph {\bibinfo {title} {Signal {Processing}: {A}
  {Mathematical} {Approach}}}}\ (\bibinfo  {publisher} {A K Peters/CRC Press},\
  \bibinfo {address} {Wellesley, Mass},\ \bibinfo {year} {1993})\BibitemShut
  {NoStop}%
\bibitem [{Note1()}]{Note1}%
  \BibitemOpen
  \bibinfo {note} {The modified Hamiltonian is defined using a partial time
  derivative and therefore neglects the implicit time-dependence of $\lambda
  _t$.}\BibitemShut {Stop}%
\bibitem [{\citenamefont {Breuer}\ and\ \citenamefont
  {Petruccione}(2007)}]{breuer_theory_2007}%
  \BibitemOpen
  \bibfield  {author} {\bibinfo {author} {\bibfnamefont {H.-P.}\ \bibnamefont
  {Breuer}}\ and\ \bibinfo {author} {\bibfnamefont {F.}~\bibnamefont
  {Petruccione}},\ }\href {\doibase 10.1093/acprof:oso/9780199213900.001.0001}
  {\emph {\bibinfo {title} {The {Theory} of {Open} {Quantum} {Systems}}}}\
  (\bibinfo  {publisher} {Oxford University Press},\ \bibinfo {year}
  {2007})\BibitemShut {NoStop}%
\bibitem [{\citenamefont {Alicki}\ and\ \citenamefont
  {Lendi}(2007)}]{alicki_quantum_2007}%
  \BibitemOpen
  \bibfield  {author} {\bibinfo {author} {\bibfnamefont {R.}~\bibnamefont
  {Alicki}}\ and\ \bibinfo {author} {\bibfnamefont {K.}~\bibnamefont {Lendi}},\
  }\href@noop {} {\bibfield  {journal} {\bibinfo  {journal} {Lecture Notes in
  Physics}\ }\textbf {\bibinfo {volume} {286}} (\bibinfo {year}
  {2007})}\BibitemShut {NoStop}%
\bibitem [{\citenamefont {Blum}(2012)}]{blum_density_2012}%
  \BibitemOpen
  \bibfield  {author} {\bibinfo {author} {\bibfnamefont {K.}~\bibnamefont
  {Blum}},\ }\href@noop {} {\emph {\bibinfo {title} {Density {Matrix} {Theory}
  and {Applications}}}}\ (\bibinfo  {publisher} {Springer Science \& Business
  Media},\ \bibinfo {year} {2012})\BibitemShut {NoStop}%
\bibitem [{\citenamefont {Wang}\ and\ \citenamefont
  {Plenio}(2016)}]{wang_necessary_2016}%
  \BibitemOpen
  \bibfield  {author} {\bibinfo {author} {\bibfnamefont {Z.-Y.}\ \bibnamefont
  {Wang}}\ and\ \bibinfo {author} {\bibfnamefont {M.~B.}\ \bibnamefont
  {Plenio}},\ }\href {\doibase 10.1103/PhysRevA.93.052107} {\bibfield
  {journal} {\bibinfo  {journal} {Physical Review A}\ }\textbf {\bibinfo
  {volume} {93}},\ \bibinfo {pages} {052107} (\bibinfo {year} {2016})},\
  \bibinfo {note} {publisher: American Physical Society}\BibitemShut {NoStop}%
\bibitem [{\citenamefont {Xu}\ \emph {et~al.}(2019)\citenamefont {Xu},
  \citenamefont {Xie}, \citenamefont {Shi}, \citenamefont {Wang}, \citenamefont
  {Xu}, \citenamefont {Wang}, \citenamefont {Wang}, \citenamefont {Plenio},\
  and\ \citenamefont {Du}}]{xu_breaking_2019}%
  \BibitemOpen
  \bibfield  {author} {\bibinfo {author} {\bibfnamefont {K.}~\bibnamefont
  {Xu}}, \bibinfo {author} {\bibfnamefont {T.}~\bibnamefont {Xie}}, \bibinfo
  {author} {\bibfnamefont {F.}~\bibnamefont {Shi}}, \bibinfo {author}
  {\bibfnamefont {Z.-Y.}\ \bibnamefont {Wang}}, \bibinfo {author}
  {\bibfnamefont {X.}~\bibnamefont {Xu}}, \bibinfo {author} {\bibfnamefont
  {P.}~\bibnamefont {Wang}}, \bibinfo {author} {\bibfnamefont {Y.}~\bibnamefont
  {Wang}}, \bibinfo {author} {\bibfnamefont {M.~B.}\ \bibnamefont {Plenio}}, \
  and\ \bibinfo {author} {\bibfnamefont {J.}~\bibnamefont {Du}},\ }\href
  {\doibase 10.1126/sciadv.aax3800} {\bibfield  {journal} {\bibinfo  {journal}
  {Science Advances}\ }\textbf {\bibinfo {volume} {5}},\ \bibinfo {pages}
  {eaax3800} (\bibinfo {year} {2019})},\ \bibinfo {note} {publisher: American
  Association for the Advancement of Science Section: Research
  Article}\BibitemShut {NoStop}%
\bibitem [{\citenamefont {Guérin}\ \emph {et~al.}(1998)\citenamefont
  {Guérin}, \citenamefont {Yatsenko}, \citenamefont {Halfmann}, \citenamefont
  {Shore},\ and\ \citenamefont {Bergmann}}]{guerin_stimulated_1998}%
  \BibitemOpen
  \bibfield  {author} {\bibinfo {author} {\bibfnamefont {S.}~\bibnamefont
  {Guérin}}, \bibinfo {author} {\bibfnamefont {L.~P.}\ \bibnamefont
  {Yatsenko}}, \bibinfo {author} {\bibfnamefont {T.}~\bibnamefont {Halfmann}},
  \bibinfo {author} {\bibfnamefont {B.~W.}\ \bibnamefont {Shore}}, \ and\
  \bibinfo {author} {\bibfnamefont {K.}~\bibnamefont {Bergmann}},\ }\href
  {\doibase 10.1103/PhysRevA.58.4691} {\bibfield  {journal} {\bibinfo
  {journal} {Physical Review A}\ }\textbf {\bibinfo {volume} {58}},\ \bibinfo
  {pages} {4691} (\bibinfo {year} {1998})},\ \bibinfo {note} {publisher:
  American Physical Society}\BibitemShut {NoStop}%
\bibitem [{\citenamefont {Yatsenko}\ \emph {et~al.}(1998)\citenamefont
  {Yatsenko}, \citenamefont {Guérin}, \citenamefont {Halfmann}, \citenamefont
  {Böhmer}, \citenamefont {Shore},\ and\ \citenamefont
  {Bergmann}}]{yatsenko_stimulated_1998}%
  \BibitemOpen
  \bibfield  {author} {\bibinfo {author} {\bibfnamefont {L.~P.}\ \bibnamefont
  {Yatsenko}}, \bibinfo {author} {\bibfnamefont {S.}~\bibnamefont {Guérin}},
  \bibinfo {author} {\bibfnamefont {T.}~\bibnamefont {Halfmann}}, \bibinfo
  {author} {\bibfnamefont {K.}~\bibnamefont {Böhmer}}, \bibinfo {author}
  {\bibfnamefont {B.~W.}\ \bibnamefont {Shore}}, \ and\ \bibinfo {author}
  {\bibfnamefont {K.}~\bibnamefont {Bergmann}},\ }\href {\doibase
  10.1103/PhysRevA.58.4683} {\bibfield  {journal} {\bibinfo  {journal}
  {Physical Review A}\ }\textbf {\bibinfo {volume} {58}},\ \bibinfo {pages}
  {4683} (\bibinfo {year} {1998})},\ \bibinfo {note} {publisher: American
  Physical Society}\BibitemShut {NoStop}%
\bibitem [{\citenamefont {Engel}\ \emph {et~al.}(2007)\citenamefont {Engel},
  \citenamefont {Calhoun}, \citenamefont {Read}, \citenamefont {Ahn},
  \citenamefont {Mančal}, \citenamefont {Cheng}, \citenamefont {Blankenship},\
  and\ \citenamefont {Fleming}}]{engel_evidence_2007}%
  \BibitemOpen
  \bibfield  {author} {\bibinfo {author} {\bibfnamefont {G.~S.}\ \bibnamefont
  {Engel}}, \bibinfo {author} {\bibfnamefont {T.~R.}\ \bibnamefont {Calhoun}},
  \bibinfo {author} {\bibfnamefont {E.~L.}\ \bibnamefont {Read}}, \bibinfo
  {author} {\bibfnamefont {T.-K.}\ \bibnamefont {Ahn}}, \bibinfo {author}
  {\bibfnamefont {T.}~\bibnamefont {Mančal}}, \bibinfo {author} {\bibfnamefont
  {Y.-C.}\ \bibnamefont {Cheng}}, \bibinfo {author} {\bibfnamefont {R.~E.}\
  \bibnamefont {Blankenship}}, \ and\ \bibinfo {author} {\bibfnamefont {G.~R.}\
  \bibnamefont {Fleming}},\ }\href {\doibase 10.1038/nature05678} {\bibfield
  {journal} {\bibinfo  {journal} {Nature}\ }\textbf {\bibinfo {volume} {446}},\
  \bibinfo {pages} {782} (\bibinfo {year} {2007})}\BibitemShut {NoStop}%
\bibitem [{\citenamefont {Zhang}\ \emph {et~al.}(2015)\citenamefont {Zhang},
  \citenamefont {Oh}, \citenamefont {Alharbi}, \citenamefont {Engel},\ and\
  \citenamefont {Kais}}]{zhang_delocalized_2015}%
  \BibitemOpen
  \bibfield  {author} {\bibinfo {author} {\bibfnamefont {Y.}~\bibnamefont
  {Zhang}}, \bibinfo {author} {\bibfnamefont {S.}~\bibnamefont {Oh}}, \bibinfo
  {author} {\bibfnamefont {F.~H.}\ \bibnamefont {Alharbi}}, \bibinfo {author}
  {\bibfnamefont {G.~S.}\ \bibnamefont {Engel}}, \ and\ \bibinfo {author}
  {\bibfnamefont {S.}~\bibnamefont {Kais}},\ }\href {\doibase
  10.1039/C4CP05310A} {\bibfield  {journal} {\bibinfo  {journal} {Physical
  Chemistry Chemical Physics}\ }\textbf {\bibinfo {volume} {17}},\ \bibinfo
  {pages} {5743} (\bibinfo {year} {2015})}\BibitemShut {NoStop}%
\bibitem [{\citenamefont {Ishizaki}\ \emph {et~al.}(2010)\citenamefont
  {Ishizaki}, \citenamefont {Calhoun}, \citenamefont {Schlau-Cohen},\ and\
  \citenamefont {Fleming}}]{ishizaki_quantum_2010}%
  \BibitemOpen
  \bibfield  {author} {\bibinfo {author} {\bibfnamefont {A.}~\bibnamefont
  {Ishizaki}}, \bibinfo {author} {\bibfnamefont {T.~R.}\ \bibnamefont
  {Calhoun}}, \bibinfo {author} {\bibfnamefont {G.~S.}\ \bibnamefont
  {Schlau-Cohen}}, \ and\ \bibinfo {author} {\bibfnamefont {G.~R.}\
  \bibnamefont {Fleming}},\ }\href {\doibase 10.1039/C003389H} {\bibfield
  {journal} {\bibinfo  {journal} {Physical Chemistry Chemical Physics}\
  }\textbf {\bibinfo {volume} {12}},\ \bibinfo {pages} {7319} (\bibinfo {year}
  {2010})}\BibitemShut {NoStop}%
\bibitem [{\citenamefont {Johnson}\ \emph {et~al.}(2015)\citenamefont
  {Johnson}, \citenamefont {Halpin}, \citenamefont {Morizumi}, \citenamefont
  {Prokhorenko}, \citenamefont {Ernst},\ and\ \citenamefont
  {Miller}}]{johnson_local_2015}%
  \BibitemOpen
  \bibfield  {author} {\bibinfo {author} {\bibfnamefont {P.~J.~M.}\
  \bibnamefont {Johnson}}, \bibinfo {author} {\bibfnamefont {A.}~\bibnamefont
  {Halpin}}, \bibinfo {author} {\bibfnamefont {T.}~\bibnamefont {Morizumi}},
  \bibinfo {author} {\bibfnamefont {V.~I.}\ \bibnamefont {Prokhorenko}},
  \bibinfo {author} {\bibfnamefont {O.~P.}\ \bibnamefont {Ernst}}, \ and\
  \bibinfo {author} {\bibfnamefont {R.~J.~D.}\ \bibnamefont {Miller}},\ }\href
  {\doibase 10.1038/nchem.2398} {\bibfield  {journal} {\bibinfo  {journal}
  {Nature Chemistry}\ }\textbf {\bibinfo {volume} {7}},\ \bibinfo {pages} {980}
  (\bibinfo {year} {2015})}\BibitemShut {NoStop}%
\bibitem [{\citenamefont {Jiang}\ and\ \citenamefont
  {Brumer}(1991)}]{jiang_creation_1991}%
  \BibitemOpen
  \bibfield  {author} {\bibinfo {author} {\bibfnamefont {X.-P.}\ \bibnamefont
  {Jiang}}\ and\ \bibinfo {author} {\bibfnamefont {P.}~\bibnamefont {Brumer}},\
  }\href {\doibase 10.1063/1.460467} {\bibfield  {journal} {\bibinfo  {journal}
  {The Journal of Chemical Physics}\ }\textbf {\bibinfo {volume} {94}},\
  \bibinfo {pages} {5833} (\bibinfo {year} {1991})}\BibitemShut {NoStop}%
\bibitem [{\citenamefont {Brumer}(2018)}]{brumer_shedding_2018}%
  \BibitemOpen
  \bibfield  {author} {\bibinfo {author} {\bibfnamefont {P.}~\bibnamefont
  {Brumer}},\ }\href {\doibase 10.1021/acs.jpclett.8b00874} {\bibfield
  {journal} {\bibinfo  {journal} {The Journal of Physical Chemistry Letters}\
  }\textbf {\bibinfo {volume} {9}},\ \bibinfo {pages} {2946} (\bibinfo {year}
  {2018})}\BibitemShut {NoStop}%
\bibitem [{\citenamefont {Koyu}\ \emph {et~al.}(2020)\citenamefont {Koyu},
  \citenamefont {Dodin}, \citenamefont {Brumer},\ and\ \citenamefont
  {Tscherbul}}]{koyu_steady-state_2020}%
  \BibitemOpen
  \bibfield  {author} {\bibinfo {author} {\bibfnamefont {S.}~\bibnamefont
  {Koyu}}, \bibinfo {author} {\bibfnamefont {A.}~\bibnamefont {Dodin}},
  \bibinfo {author} {\bibfnamefont {P.}~\bibnamefont {Brumer}}, \ and\ \bibinfo
  {author} {\bibfnamefont {T.~V.}\ \bibnamefont {Tscherbul}},\ }\href@noop {}
  {\bibfield  {journal} {\bibinfo  {journal} {arXiv preprint arXiv:2001.09230}\
  } (\bibinfo {year} {2020})}\BibitemShut {NoStop}%
\bibitem [{\citenamefont {Reppert}\ \emph {et~al.}(2019)\citenamefont
  {Reppert}, \citenamefont {Reppert}, \citenamefont {Pachon},\ and\
  \citenamefont {Brumer}}]{reppert_equilibrium_2019}%
  \BibitemOpen
  \bibfield  {author} {\bibinfo {author} {\bibfnamefont {M.}~\bibnamefont
  {Reppert}}, \bibinfo {author} {\bibfnamefont {D.}~\bibnamefont {Reppert}},
  \bibinfo {author} {\bibfnamefont {L.~A.}\ \bibnamefont {Pachon}}, \ and\
  \bibinfo {author} {\bibfnamefont {P.}~\bibnamefont {Brumer}},\ }\href
  {http://arxiv.org/abs/1911.07606} {\bibfield  {journal} {\bibinfo  {journal}
  {arXiv:1911.07606 [quant-ph]}\ } (\bibinfo {year} {2019})},\ \bibinfo {note}
  {arXiv: 1911.07606}\BibitemShut {NoStop}%
\bibitem [{\citenamefont {Axelrod}\ and\ \citenamefont
  {Brumer}(2018)}]{axelrod_efficient_2018}%
  \BibitemOpen
  \bibfield  {author} {\bibinfo {author} {\bibfnamefont {S.}~\bibnamefont
  {Axelrod}}\ and\ \bibinfo {author} {\bibfnamefont {P.}~\bibnamefont
  {Brumer}},\ }\href {\doibase 10.1063/1.5041005} {\bibfield  {journal}
  {\bibinfo  {journal} {The Journal of Chemical Physics}\ }\textbf {\bibinfo
  {volume} {149}},\ \bibinfo {pages} {114104} (\bibinfo {year} {2018})},\
  \bibinfo {note} {publisher: American Institute of Physics}\BibitemShut
  {NoStop}%
\bibitem [{\citenamefont {Axelrod}\ and\ \citenamefont
  {Brumer}(2019)}]{axelrod_multiple_2019}%
  \BibitemOpen
  \bibfield  {author} {\bibinfo {author} {\bibfnamefont {S.}~\bibnamefont
  {Axelrod}}\ and\ \bibinfo {author} {\bibfnamefont {P.}~\bibnamefont
  {Brumer}},\ }\href {\doibase 10.1063/1.5099969@jcp.2019.OSQD2019.issue-1}
  {\bibfield  {journal} {\bibinfo  {journal} {The Journal of Chemical Physics}\
  }\textbf {\bibinfo {volume} {OSQD2019}},\ \bibinfo {pages} {014104} (\bibinfo
  {year} {2019})},\ \bibinfo {note} {publisher: American Institute of
  Physics}\BibitemShut {NoStop}%
\bibitem [{\citenamefont {Tscherbul}\ and\ \citenamefont
  {Brumer}(2014)}]{tscherbul_long-lived_2014}%
  \BibitemOpen
  \bibfield  {author} {\bibinfo {author} {\bibfnamefont {T.~V.}\ \bibnamefont
  {Tscherbul}}\ and\ \bibinfo {author} {\bibfnamefont {P.}~\bibnamefont
  {Brumer}},\ }\href {\doibase 10.1103/PhysRevLett.113.113601} {\bibfield
  {journal} {\bibinfo  {journal} {Physical Review Letters}\ }\textbf {\bibinfo
  {volume} {113}},\ \bibinfo {pages} {113601} (\bibinfo {year}
  {2014})}\BibitemShut {NoStop}%
\end{thebibliography}%

\end{document}